\documentclass[journal,twoside]{IEEEtran}

\usepackage{amsmath}
\usepackage{amssymb}
\usepackage{latexsym}
\usepackage{graphicx}
\usepackage{epsfig}
\usepackage{psfrag}
\usepackage[active]{srcltx} 
\usepackage{color}
\usepackage{setspace}
\usepackage[normal,footnotesize]{caption}
\usepackage{balance}
\usepackage{textcomp}
\usepackage{url}
\usepackage{subfig}

\newcommand{\insertonefig}[4]
{
\begin{figure}[htbp]
\begin{center}
\includegraphics[width=#1,clip,keepaspectratio]{#2}
\end{center}
\caption{#4}
{#3}
\end{figure}
}

\newcommand{\addphoto}[1]{\includegraphics[width=1in,height=1.25in,clip,keepaspectratio]{#1}}

\newcommand{\noi} {\noindent}

\begin{document}


\title{For the Grid and Through the Grid:\\
The Role of Power Line Communications\\in the Smart Grid}

\author{Stefano Galli\IEEEmembership{,~Senior~Member,~IEEE,}
        Anna Scaglione\IEEEmembership{,~Fellow,~IEEE,}
        Zhifang Wang\IEEEmembership{,~Member,~IEEE}

\thanks{\noi Stefano Galli (sgalli@assia-inc.com) is now with ASSIA, Inc. At the time of writing this paper, Dr. Galli was with Panasonic Corporation.}

\thanks{Anna Scaglione (ascaglione@ucdavis.edu) is with University of California at Davis, Davis, CA, USA.}
\thanks{Zhifang Wang (zfwang@illinois.edu) is with University of Illinois at Urbana-Champaign, Urbana, IL, USA.}

\thanks{A. Scaglione and Z. Wang acknowledge that their work was funded in part by the UIUC TCIP Project under the Grant NSF CNS 05­24695, and in part by the TCIPG Project  which is sponsored by the Department of Energy under
Award Number DE-OE0000097.}%

\thanks{Submitted May 28,~2010. Revised Sep. 12, 2010, and Jan. 12, 2011.}%

}

\markboth{To appear in the Proceedings of the IEEE - June 2011}%
{S. Galli \MakeLowercase{\textit{et al.}}: For the Grid and Through the Grid: The Role of PLC in the Smart Grid}


\maketitle

\begin{abstract}
Is Power Line Communications (PLC) a good candidate for Smart Grid applications?
The objective of this paper is to address this important question. To do so we provide an overview of what PLC can deliver today by surveying its history and describing the most recent technological advances in the area. We then address Smart Grid applications as instances of sensor networking and network control problems and discuss the main conclusion one can draw from the literature on these subjects. The application scenario of PLC within the Smart Grid is then analyzed in detail.  Since a necessary ingredient of network planning is modeling, we also discuss two aspects of engineering modeling that relate to our question. The first aspect is modeling the PLC channel through fading models. The second aspect we review is the Smart Grid control and traffic modeling problem which allows us to achieve a better understanding of the communications requirements. Finally, this paper reports recent studies on the electrical and topological properties of a sample power distribution network. Power grid topological studies are very important for PLC networking as the power grid is not only the information source \textit{but also} the  information delivery system - a unique feature when PLC is used for the Smart Grid.
\end{abstract}


\thispagestyle{empty}

\section{Introduction}
Communications over power lines (PLs) is an old idea that dates back to the early 1900's, when the first patents were filed in this area \cite{Schwartz2008historyPLC}. Since then, utility companies around the world have been using this technology for remote metering and load control \cite{Book:Dostert2001, Book:PLC-NEW-2010}, using at first single carrier narrowband (NB) solutions operating in the Audio/Low Frequency bands that achieved data rates ranging from few bps to a few kbps. As technology matured and the application space widened, broadband (BB) PLC systems operating in the High Frequency band (2-30 MHz) and achieving data rates up to a 200 Mbps started to appear in the market. In the last few years, industry interest has also grown around the so-called ``high data rate'' NB-PLC based on multicarrier schemes and operating in the band between 3-500 kHz.

PLC is also used to provide BB Internet access to residential customers, BB LAN connectivity within home/office/vehicles, and command and control capabilities for automation and remote metering \cite{LatYon2003, PavHanYaz2003, GalScaDos2003, BigGalLee2006}. The basic incentive for using PLC is that the power grid provides an infrastructure that is much more extensive and pervasive than any other wired/wireless alternative, so that virtually every line-powered device can become the target of value-added services.

In spite of the PLC promise of being the enabler of a multitude of present and future applications, PLC has not yet reached the mass market penetration that is within its potential.
However, a new compelling reason for using PLC is today emerging: the recent impetus in modernizing the aging power grid through an information highway dedicated to the management of energy transmission and distribution, the so called {\it Smart Grid}. It is commonly recognized that the Smart Grid will be supported by a heterogeneous set of networking technologies, as no single solution fits all scenarios. Nevertheless, an interesting question is whether the Smart Grid will have a pivotal role in fostering the success of PLC in the market.

The objective of this paper is to analyze critically the role of sensing, communications, and control in the Smart Grid and, at the same time, clarify what PLC can offer today and what is unique to PLC for Smart Grid applications.

\subsection{The Smart Grid Design Challenge}
It is broadly believed that the growth of energy demand has outpaced the rate at which energy generation can grow by traditional means. Additionally, many governments agree that greenhouse gas emissions need to be contained to control or prevent climate change. The necessity of modernizing the electric grid infrastructure around the world is both the consequence of the limited investments made in it in the last decades, as well as of the result of new requirements that emerge in the safe integration of utility scale Renewable Energy Sources feeding into the transmission system, Distributed Energy Resources (DER) feeding into the distribution system or the home, decentralized storage to compensate for the time varying nature of wind and photovoltaic sources, Plug-in (Hybrid) Electric Vehicles (PHEV) that may cause large load increases on sections of the grid, microgrids, and in allowing active participation of consumers via Demand Side Management (DSM) and Demand Response (DR) programs - all of which are advocated as sustainable solutions to our energy crisis \cite{HeydLiuPhad01}.

Balancing generation and demand at a very granular scale requires the integration of additional protection and control technologies that ensure grid stability \cite{Garrity:2008, IpakAlbu2009} and that are not a trivial patch to the
current distribution network \cite{WuMoslBose2005}. Power grids are designed to be managed through a rather old-fashioned centralized cyber-infrastructure model, referred to as Supervisory Control and Data Acquisition (SCADA). Hence, the concept of {\it Smart Grid} has emerged, encompassing the cyber-physical infrastructure including wide-area monitoring, two way communications and enhanced control functionalities that will bridge the present technological inadequacies of the SCADA system.

Since communications is such a fundamental element of the Smart Grid, the appropriate design for physical, data and network communications layers are today a topic of intense debate. Unfortunately, Smart Grid is today more a ``vision'' than an actual design. Quoting Tomsovic et al.: ``\textit{Although the available communication today is fast enough, the computation needed for such real-time control is still very complex and poorly understood}''~\cite{TomsBakkVenkBose2005}. For instance, DR and load shedding can potentially yield economic benefits and energy savings \cite{HoppGoldBhar2007SummerDR, FERC:DemandResponse:2009}, however the correct implementation of DR and, more in general, of DSM applications for maximizing system savings under stability constraints is still not known. For example, in \cite{ZemaProkGuo08} it is pointed out that: ``\textit{When demand peaks occur, reducing energy to a minimum seems like a valid solution, however this compromises system stability.}''

Simulations and small field trials - often conducted with cautious containment to prevent cascading failures - are insufficient to grasp fundamental threats to the global stability of the grid that can arise when dealing with a large scale system. In fact, still absent in most technical discussions are specific parameters of the monitoring and control functions that Smart Grid communications shall enable. Furthermore, the optimality at the level of sub-systems is no guarantee of an overall optimal design. As we discuss in more detail in Section \ref{sec.ControlSensorNetworking}, optimizing communications, control and sensing in a large decentralized cyber-physical system is a very complex and elusive problem. For instance, the results available on observability and stability of networked control systems are valid only under very restrictive assumptions (e.g. a single link with zero latency, a perfectly known system, etc.) \cite{IshiiFrancis2002book, Baillieul2002book, FagnaniZampieri03, TatiMitter04}.

In order to design a communications scheme and examine its efficiency from both a scalability and a distributed control point of view, it is of paramount importance to characterize statistically the Smart Grid information \textit{source}, i.e. the power grid itself. As for any interconnected system, the dynamics sensed are highly coupled and dependent, an aspect that should not be ignored in managing, aggregating and prioritizing the network traffic. However, very little work has been done in this direction. Interestingly, in the case of PLC, the characterization of the grid as an information source will also lead to a better understanding of the grid as an information delivery system since the grid becomes \textit{also} the physical network medium when PLC is used.

\subsection{PLC and the Smart Grid}
The debate on what is the actual role of PLC in the Smart Grid is still open and ongoing: while some advocate that PLC is a very good candidate for many applications, others express concerns and look at wireless as a more established alternative. There is no doubt that the Smart Grid will exploit multiple types of communications technologies, ranging from fiber optics to wireless and to wireline. Skeptics contend that PLC has an unclear standardization status and offers data rates that are too small; others also contend that PLC modems are still too expensive and that they present electromagnetic compatibility (EMC) issues.  Recent advances in PLC clear much of these concerns.

Among the wireline alternatives, PLC is the only technology that has deployment cost that can be considered comparable to wireless since the lines are already there. A promising sign, confirming that PLC has already exited the experimental phase and is a technology mature for deployment, is the extensive use of PLC over both the transmission and the distribution parts of the grid - including the wide penetration that PLC has gained for supporting Automatic Meter Reading (AMR) and Advanced Metering Infrastructure (AMI) applications.

This said, there are two aspects that could hinder the success of PLC in the market. One is the commercial pressure to jump on the bandwagon of Smart Grid applications with the wrong PLC technology. Especially in the US, some PLC vendors are promoting for Smart Grid applications the exclusive use of BB-PLC modems which were not designed for Smart Grid applications but were originally designed to support Home/Building Area Networks (HAN/BAN) or Internet access applications. These solutions have limited range and are likely to be over-designed for Smart Grid applications. Furthermore, as we will show in Sect. \ref{sec.PLCforSmartgrid}, there are several PLC technologies that can be used in Smart Grid applications and advocating the use of a single class of PLC technologies is not a sensible approach compared to choosing the right PLC technology for the right application. A second impediment for the adoption of PLC in the Smart Grid is due to the PLC standardization status. In the last couple of years, the PLC industry has moved from a complete lack of BB-PLC standards to the opposite extreme of having multiple non-interoperable technologies ratified by several Standards Developing Organizations (SDOs), i.e. TIA-1113, ITU-T G.hn, IEEE 1901 FFT-OFDM, and IEEE 1901 Wavelet-OFDM \cite{Book:PLC-NEW-2010Chap7}. Thus, advocating the exclusive use of BB-PLC modems for Smart Grid applications, raises the issue of which standard to use as well as the issue of their incompatibility. A similar fate on the availability of multiple non-interoperable technologies may be expected for multicarrier HDR NB-PLC being today standardized in IEEE P1901.2 and ITU-T G.hnem. This situation leads to confusion in the market and deployments are delayed.

Interference between non-interoperable devices is the likely side effect of today's industry fragmentation. This problem has been somewhat overlooked in Smart Grid recommendations which \textit{implicitly} assume that interference is manageable or absent. Fortunately, there are today standardized mechanisms that limit the harmful interference caused by non-interoperable neighboring devices; these mechanisms are commonly referred to as ``coexistence mechanisms'' - see Sect. \ref{sec.STDscoexistence} for more details. A coexistence success story is the PHY/MAC-agnostic coexistence scheme CENELEC EN 50065 \cite{STD:Cenelec1992} which has not only allowed the ratification of several NB-PLC standards after its publication in 1992 but has also allowed NB-PLC solutions to flourish in the market for the last two decades. On the other hand, there are also coexistence-skeptics who believe that coexistence will foster the proliferation of non-standardized solution that, in turn, will cause a delay in aligning the market behind a single PLC standard. As a consequence, this market confusion and uncertainty has further delayed the adoption of PLC, especially in the US. The issue of coexistence has been addressed in detail in the Priority Action Plan 15 (PAP 15) instituted by the US National Institute of Standards and Technology (NIST) to tackle issues that would prevent the correct functioning of PLC-based Smart Grid applications \cite{Link:NIST-PAP-15}. This group has recently given a clear and strong answer to this matter by issuing a set of recommendations that require, among other things, to \textit{mandate} the support of coexistence in all BB-PLC implementations - for the full report, see \cite{TR:PAP15-BB-Recs}. Hopefully, this clear recommendation will lead to more clarity in the industry.

In the following, we will mainly refer to three classes of PLC technologies\footnote{$^)$~BB-PLC technologies devoted to Internet-access applications have also been referred to as Broadband over Power Lines or BPL, whereas LDR NB-PLC technologies have been referred to as Distribution Line Carrier or Power Line Carrier.}:

\textit{Ultra Narrow Band (UNB)}: Technologies operating at very low data rate (\texttildelow 100 bps) in the Ultra Low Frequency (0.3-3 kHz) band or in the upper part of the Super Low Frequency (30-300 Hz) band. An historical example of a one-way communication link supporting load control applications is Ripple Carrier Signaling which operates in the 125 - 2,000 kHz and is able to convey several bps band using simple Amplitude Shift Keying modulation. More recent examples are the AMR Turtle System which conveys data at extremely low speed (\texttildelow 0.001 bps) and the Two-Way Automatic Communications System (TWACS) that can carry data at a maximum data rate of two bits per mains frequency cycle, i.e. 100 bps in Europe and 120 bps in North America. UNB-PLC have a very large operational range (150 km or more). Although the data rate per link is low, deployed systems use various forms of parallelization and efficient addressing that support good scalability capabilities. Despite the fact that these UNB solutions are proprietary, they are very mature technologies, they have been in the field for at least two decades, and have been deployed by hundreds of utilities.

\textit{Narrowband (NB)}: Technologies operating in the VLF/LF/MF bands (3-500 kHz), which include the European CENELEC (Comit\'{e} Europ\'{e}en de Normalisation \'{E}lectrotechnique) bands (3-148.5 kHz), the US FCC (Federal Communications Commission) band (10-490 kHz), the Japanese ARIB (Association of Radio Industries and Businesses) band (10-450 kHz), and the Chinese band (3-500 kHz). Specifically, we have:
        \begin{itemize}
              \item \textit{Low Data Rate (LDR)}: Single carrier technologies capable of data rates of few kbps. Typical examples of LDR NB-PLC technologies are devices conforming to the following recommendations: ISO/IEC 14908-3 (LonWorks), ISO/IEC 14543-3-5 (KNX), CEA-600.31 (CEBus), IEC 61334-3-1, IEC 61334-5 (FSK and Spread-FSK), etc. Additional non-SDO based examples are Insteon, X10, and HomePlug C\&C, SITRED, Ariane Controls, BacNet etc.

              \item \textit{High Data Rate (HDR)}: Multicarrier technologies capable of data rates ranging between tens of kbps and up to 500 kbps. Typical examples of HDR NB-PLC technologies are those devices within the scope of ongoing standards projects: ITU-T G.hnem, IEEE 1901.2. Additional non-SDO based examples are PRIME and G3-PLC.
        \end{itemize}

\textit{Broadband (BB)}: Technologies operating in the HF/VHF bands (1.8-250 MHz) and having a PHY rate ranging from several Mbps to several hundred Mbps. Typical examples of BB-PLC technologies are devices conforming to the TIA-1113 (HomePlug 1.0), IEEE 1901, ITU-T G.hn (G.9960/G.9961) recommendations. Additional non-SDO based examples are HomePlug AV/Extended, HomePlug Green PHY, HD-PLC, UPA Powermax, and Gigle MediaXtreme.

\subsection{Organization of Work}
This paper\footnote{$^)$~Initial results have been presented at the \textit{First IEEE International Conference on Smart Grid Communications} (SmartGridComm 2010) \cite{GalliScaglioneWang10SGcomm}.} starts with a brief historical overview of PLC in Sect. \ref{sec.HistoryPLC}, and then reports on the status of the most recent PLC standards in Sect. \ref{sec.STDs}. The role of communication, sensing, and control in the Smart Grid is addressed in Sect. \ref{sec.RoleComms} by looking at the required evolution path of today's SCADA systems and highlighting the most salient issues related to control and sensor networking, as well as tackling the problem of characterizing the traffic that needs to be supported. The specific role that PLC can have in the Smart Grid is then addressed in Sect. \ref{sec.PLCforSmartgrid}, where applications to the transmission and distribution parts of the grid are analyzed. We will then dedicate Sect. \ref{sec.DeploymentAspects} to discuss fundamental design issues. Recognizing that an important element of network design is the availability of planning tools for its deployment, we will first review the state of the art in PLC channel modeling in Sect. \ref{subsec.ChannelModels}; furthermore, in Sects. \ref{sec.Zhifang} and \ref{subsec.ZhifangLV} we will make the first step at analyzing the grid as both a data source and as an information delivery system - as PLC naturally entails. Final considerations and recommendations are then made in Sect. \ref{sec.Conclusions}. The list of acronyms used throughout the paper can be found in Table \ref{tab.acronyms}.

\begin{table*}[htbp]
  \centering
  \caption{List of acronyms used in the paper.}\label{tab.acronyms}
\small
\begin{tabular}{llcll}
  \hline\hline
  Acronym & Meaning & & Acronym & Meaning \\ [0.5ex]
  \hline
AC	&	Alternate Current	&		&	ISP	&	Inter System Protocol	\\
AMI	&	Advanced Metering Infrastructure	&		&	ITU	&	International Telecommunication Union	 \\
AMR	&	Automatic Meter Reading	&		&	LDR	&	Low Data Rate	\\
ARIB	&	Association of Radio Industries and Businesses	&		&	LV	&	Low Voltage	\\
BB	&	Broad Band	&		&	MAC	&	Medium Access Control	\\
CDMA	&	Code Division Multiple Access	&		&	MoCA	&	Multimedia over Coax Alliance	\\
CENELEC	&	Comit\'{e} Europ\'{e}en de Normalisation \'{E}lectrotechnique	&		&	MTL	&	 Multi-Conductor Transmission Line	 \\
CEPCA	&	Consumer Electronics Powerline Alliance	&		&	MV	&	Medium Voltage	\\
CSMA/CA	&	Carrier Sense Multiple Access/Collision Avoidance	&		&	NB	&	Narrowband	\\
DC	&	Direct Current	&		&	NIST	&	National Institute of Standards and Technology	\\
DER	&	Distributed Energy Resources	&		&	OFDM	&	Orthogonal  Frequency Division Multiplexing  	\\
DR	&	Demand Response	&		&	PAP	&	Priority Action Plan	\\
DSM	&	Demand Side Management	&		&	PHEV	&	Plug-in (Hybrid) Electric Vehicles	\\
EVSE	&	Electric Vehicle Supply Equipment	&		&	PHY	&	Physical Layer	\\
FACTS	&	Flexible AC Transmission System	&		&	PL	&	Power Line	\\
FCC	&	Federal Communications Commission	&		&	PLC	&	Power Line Communications	\\
FDM	&	Frequency Division Multiplexing	&		&	PMU	&	Phasor Measurement Unit	\\
FSK	&	Frequency Shift Keying	&		&	PRIME	&	Powerline Related Intelligent Metering Evolution	\\
HAN	&	Home Area Network	&		&	REMPLI	&	Real-time Energy Management via Power lines and Internet	\\
HD-PLC	&	High Definition Power Line Communication	&		&	RTU	&	Remote Terminal Unit	\\
HDR	&	High Data Rate	&		&	SAE	&	Society of Automotive Engineers	\\
HEMS/BEMS	&	Home/Building Energy Management System	&		&	SCADA	&	Supervisory Control and Data Acquisition	 \\
HomeGrid	&	HomeGrid Forum	&		&	SDO	&	Standard Development Organization	\\
HomePlug	&	HomePlug Powerline Alliance	&		&	TDMA	&	Time Division Multiple Access	\\
HomePNA	&	Home Phone Networking Alliance	&		&	TIA	&	Telecommunications Industry Association	 \\
HV	&	High Voltage	&		&	TL	&	Transmission Line	\\
IEC	&	International Electrotechnical Commission	&		&	TWACS	&	Two-Way Automatic Communications System	 \\
IED	&	Intelligent Electronic Devices	&		&	UNB	&	Ultra Narrowband	\\
IEEE	&	Institute of Electrical and Electronics Engineers	&		&	UPA	&	Universal Powerline Association	 \\
IPP	&	Inter-PHY Protocol	&		&	WAMS	&	Wide Area Measurement System	\\
ISO	&	International Organization for Standardization	&		&	WSCC	&	Western Systems Coordinating Council 	 \\[1ex]

\hline \hline
\end{tabular}
\end{table*}

\section{Historical Overview of PLC} \label{sec.HistoryPLC}

\subsection{The Early Years}
The first PLC applications put in place by power utilities involved voice and data communications over High Voltage (HV) lines which typically bear voltages above 100 kV and span very large geographical distances. HV lines have been used as a communications medium for voice since the 1920s (power carrier systems) \cite{Schwartz2008historyPLC}. In those years telephone coverage was very poor and engineers operating power plants and transformer stations used PLC as an alternative way to communicate for operations management with colleagues stationed tens or hundreds of km away.

When digital communications techniques were later introduced, only very low data rates (few hundred bps) were achievable for supporting telemetering and tele-control applications \cite{Book:Dostert2001, Book:PLC-NEW-2010}.

Another important driver for the original interest of utilities in PLC was load control, i.e. the capability of switching on/off appliances responsible for high energy consumption such as air conditioners, water heaters, etc. Utilities have been using Ripple Carrier Signaling since the 1930s to control peak events at demand side by issuing control signals to switch off heavy duty appliances \cite{Book:Dostert2001}.
Ripple Carrier Signaling has been quite successful, especially in Europe, and its use has been extended to include other applications such as day/night tariff switching, street light control, and control of the equipment on the power grid. Interestingly, load control is attracting renewed interest as a means to balance generation and demand - see \cite{HoppGoldBhar2007SummerDR, FERC:DemandResponse:2009} for an analysis of savings and benefits of DSM.

\subsection{Ultra Narrowband and Narrowband PLC} \label{sec.UNBandNB}
In the last couple of decades, several AMR/AMI solutions using PLC, wireless, and phone lines have been deployed by utilities. As far as PLC, first deployments involved UNB-PLC technologies like the Turtle System \cite{Nordell08amr} and TWACS \cite{MakReed1982twacs1, MakMoore1984wacs2}. Both systems use disturbances of the voltage waveform for outbound (substation to meter) communication and of the current waveform for inbound (meter to substation) communication. The Turtle System has been mostly used for AMR as the first available products (TS1) allowed only one-way inbound connectivity; a two-way version (TS2) of the Turtle System became available after 2002. On the other hand, TWACS is widely used especially in the US for AMI, distribution automation, and DR application. TWACS allows several levels of parallelization that effectively increase its capability of handling several tens or hundreds of thousands of end-points (meters): 6 TWACS channel per phase are conveyed using Code Division Multiple Access (CDMA) with orthogonal codes (Hadamard); feeders can be operated independently and simultaneously; an efficient addressing scheme allows handling small or large groups of end-points via polling. This combination of TDMA, CDMA and group polling allows TWACS to minimize the issues related to channel contention so that few hundred thousand meter readings can be carried out within an hour with the appropriate configuration \cite{PrivComm:TWACSnov2010}. A narrowband TWACS-like method that could provide an inbound data rate higher than TWACS has been recently proposed in \cite{RiekWalk10ulf}.

Recognizing the increasing desire for higher data rate, CENELEC issued in 1992 standard EN 50065 \cite{STD:Cenelec1992}. The CENELEC EN 50065 standard allows communication over Low Voltage (LV) distribution PL in the frequency range from 3 kHz up to 148.5 kHz. Four frequency bands are defined:

\begin{itemize}
\item A (3-95 kHz): reserved exclusively to power utilities.
\item B (95-125 kHz): any application.
\item C (125-140 kHz): in-home networking systems with a \textit{mandated} Carrier Sense Multiple Access with Collision Avoidance (CSMA/CA) protocol.
\item D (140-148.5 kHz): alarm and security systems.
\end{itemize}

CENELEC mandates a CSMA/CA mechanism (EN 50065) in the C-band and stations that wish to transmit must use the 132.5 kHz frequency to inform that the channel is in use \cite{STD:Cenelec1992}. This mandatory protocol defines a maximum channel holding period (1 s), a minimum pause between consecutive transmissions from the same sender (125 ms), and a minimum time for declaring the channel is idle (85 ms). Note that CENELEC specifications regulate only spectrum usage and the CSMA/CA protocol but do not mandate any modulation or coding schemes. Interestingly, the coexistence mechanism defined in \cite{STD:Cenelec1992} is PHY/MAC-agnostic and several NB-PLC standards were developed after EN 50065 was ratified.

In other countries regulations are different. For example, in the US and Asia the use of up to \texttildelow$500$ kHz is allowed by FCC and ARIB. On the other hand, FCC and ARIB have not assigned specific bands to exclusive use of the utilities so that any device can access the whole $500$ kHz and no coexistence protocol is mandated as for the CENELEC C-band.

\subsection{Broadband PLC}
As NB-PLC started to be progressively successful, BB-PLC started to appear as well - initially for Internet access applications and successively for HAN and A/V applications. The first wave of interest into the use of BB-PLC for Internet access started in Europe when Nortel and Norweb Communications in the U.K. announced in 1997 that they had developed a technology to provide access service to residential customers via PLC \cite{Clarck1998norweb}. Limited trials of broadband Internet access through PLs were conducted in Manchester and NorWeb prototypes were able to deliver data at rates around 1 Mbps. However, higher than anticipated costs and growing EMC issues caused the early termination of the project in 1999. Other projects in Europe led by Siemens and Ascom encountered a similar fate. On the other hand, a multi-year project funded by the European Community (The Open PLC European Research Alliance, OPERA) led most of the recent research efforts in the field of BB-PLC for Internet access \cite{Link:Opera}.

Given the disappointing results in using PLC for Internet access applications, the interest of industry started shifting towards in-home applications in early 2000. In the last decade, several industry alliances were formed with a charter to set technology specification mostly for in-home PLC, e.g. the HomePlug Powerline Alliance (HomePlug), Universal Powerline Association (UPA), High Definition Power Line Communication (HD-PLC) Alliance, and The HomeGrid Forum. Products allowing PHY data rates of 14 Mbps (HomePlug 1.0), then 85 Mbps (HomePlug Turbo), and then 200 Mbps (HomePlug AV, HD-PLC, UPA) have been progressively available on the market over the past several years. However, none of these technologies are interoperable with each other.

\section{The Status of PLC Standardization} \label{sec.STDs}
A comprehensive and up to date review of PLC standards can be found in \cite{Book:PLC-NEW-2010Chap7}. In the next few Sections, we will focus on the latest standardization developments that occurred in both NB and BB-PLC.

\subsection{Narrowband PLC Standards} \label{sec.STDsnarrow}
One of the first LDR NB-PLC standards ratified is the ANSI/EIA 709.1 standard, also known as LonWorks. Issued by ANSI in 1999, it became an international standard in 2008 (ISO/IEC 14908-1) \cite{STD:ISOIEC2008LonWorks}. This seven layer OSI protocol provides a set of services that allow the application program in a device to send and receive messages from other devices in the network without needing to know the topology of the network or the functions of the other devices. LonWorks transceivers are designed to operate in one of two frequency ranges depending on the end application. When configured for use in electric utility applications, the CENELEC A-band is used, whereas in-home/commercial/industrial applications use the C-band. Achievable data rates are in the order of few kbps.

The most widespread PLC technologies deployed today are based on Frequency Shift Keying (FSK) or Spread-FSK as specified in the IEC 61334-5-2 \cite{STD:IEC1998FSK} and IEC 61334-5-1 \cite{STD:IEC2001SpreadFSK} standards, respectively. The availability of standards for these technologies goes from recommendations that specify the stack of communications protocols from the physical up to the application layer (IEC 62056-53 for COSEM) thus facilitating the development of interoperable solutions. Such AMR/AMI solutions are now provided by a number of companies and are widely and successfully implemented by utilities. Spread-FSK based NB-PLC devices are being currently deployed in Europe.

There is today also a growing interest in HDR NB-PLC solutions operating in the CENELEC/FCC/ARIB bands and are able to provide higher data rates than LDR NB-PLC. For example, the recent Powerline Related Intelligent Metering Evolution (PRIME) initiative has gained industry support in Europe and has specified an HDR NB-PLC solution based on Orthogonal Frequency Division Multiplexing (OFDM), operating in the CENELEC-A band, and capable of PHY data rates up to 125 kbps \cite{Link:PRIME}. A similar initiative, G3-PLC, was also recently released \cite{Link:G3specs}. G3-PLC is an OFDM-based HDR NB-PLC specification that supports IPv6 internet-protocol standard and can operate in the $10-490$ kHz band. Recent field trials results of PRIME and G3-PLC have been reported in \cite{BergSendArz10prime} and \cite{RazaUnarKama2010g3}. Both PRIME and G3-PLC specifications are open specifications available online.

There are today two SDO-backed efforts for the standardization of HDR NB-PLC technologies, both started in early 2010: ITU-T G.hnem and IEEE 1901.2. The goal of G.hnem and P1901.2 is to define a HDR NB-PLC technology of very low complexity optimized for energy management spanning from HAN to AMI and PHEV applications and operating over both Alternate (AC) and Direct (DC) Current lines. The standards will support communications through the Medium Voltage (MV)/LV transformer, over MV lines, and over indoor and outdoor LV lines and will support data rates scalable up to 500 kbps depending on the application requirements. Within the scope of these standards there is also the design of coexistence mechanisms between HDR NB-PLC technologies and between HDR and existing LDR NB-PLC standardized technologies.

\subsection{The TIA-1113 Standard}
The world's first BB-PLC ANSI standard to be approved is the TIA-1113 \cite{STD:TIA1113:2008}. The standard is largely based on the HomePlug 1.0 specifications \cite{lee:2003} and defines a 14 Mbps PHY based on OFDM \cite{Book:PLC-NEW-2010Chap7}. Carriers are modulated with either BPSK or QPSK depending on the channel quality and operational functionality. The Media Access Control (MAC) for HomePlug 1.0 is based on a CSMA/CA scheme that features an adaptive window size management mechanism in conjunction with four levels of priority. Products based on the TIA-1113/HomePlug 1.0 specifications have experienced a good success in the in-home and industrial markets.

\subsection{The IEEE 1901 Broadband over Power Lines Standard} \label{sec.STDs1901}
The IEEE 1901 Working Group was established in 2005 to unify PL technologies with the goal of developing a standard for high-speed ($>$$100$ Mbps) communication devices using frequencies below 100 MHz and addressing both HAN and access applications \cite{Book:PLC-NEW-2010Chap7, GalLog2008, GoldTana2010access1901, RahmHongLee20XXcomparisonMAC}. The standard has been approved in September 2010 and defines two BB-PLC technologies \cite{STD:IEEE1901:2010}: an FFT-OFDM based PHY/MAC \cite{AfkhKataYong2005homeplugAV} and a Wavelet-OFDM based PHY/MAC \cite{GalKogKod2008}. Another key component of the IEEE 1901 standard is the presence of a mandatory coexistence mechanism called the Inter-System Protocol (ISP) that allows 1901-based PLC devices to share the medium fairly regardless of their PHY differences; furthermore, the ISP also allows IEEE 1901 devices to coexist with devices based on the ITU-T G.hn standard. The ISP is a new element that is unique to the PL environment - see Sect. \ref{sec.STDscoexistence}.

The FFT-OFDM IEEE 1901 PHY/MAC specification facilitates backward compatibility with devices based on the HomePlug AV specification \cite{AfkhKataYong2005homeplugAV}. Similarly, the Wavelet-OFDM IEEE 1901 PHY/MAC specification \cite{GalKogKod2008} facilitates backwards compatibility with devices based on the HD-PLC specifications of the HD-PLC Alliance led by Panasonic. The multi-PHY/MAC nature of the IEEE 1901 standard does not descend from a technical necessity but it is simply the consequence of a compromise caused by the lack of industry alignment behind a single technology. On the other hand, we can consider the multi-PHY/MAC nature of the IEEE 1901 standard as the first step towards that further consolidation of PLC technologies that will inevitably happen in the future.

Devices conforming to the standard must be capable of at least 100 Mbps and must include ISP in their implementation. Mandatory features allow IEEE 1901 devices achieving \texttildelow200 Mbps PHY data rates, while the use of optional bandwidth extending above 30 MHz allows achieving somewhat higher data rates. However, data rate improvements due to the use of higher frequencies are often marginal and characterized by short range due to the higher attenuation of the medium and the presence of TV broadcast channels above 80 MHz.

\subsection{The ITU-T G.hn Home Networking Standard} \label{sec.STDsGhn}
The ITU-T started the G.hn project in 2006 with a goal of developing a worldwide recommendation for a unified HAN transceiver capable of operating over all types of in-home wiring: phone lines, PLs, coax and Cat 5 cables and bit rates up to 1 Gbps \cite{Book:PLC-NEW-2010Chap7, OksGal2009, RahmHongLee20XXcomparisonMAC}. The PHY of G.hn was ratified by the ITU-T in October 2009 as Recommendation G.9960 \cite{STD:ITU2009GhnPHY} while the Data Link Layer was ratified in June 2010 as Recommendation G.9961 \cite{STD:ITU2010GhnMAC}. The technology targets residential houses and public places, such as small/home offices, multiple dwelling units or hotels, and does not address PLC access applications as IEEE 1901 does. Compliance to ITU-T Recommendations G.9960/G.9961 does not require the support for coexistence. Thus the support of ISP is optional for G.hn-compliant transceivers.

Past approaches emphasized transceiver optimization for a single medium only, i.e. either for PLs (HD-PLC Alliance, HomePlug, UPA), or phone lines (HomePNA), or coax cables (MoCA) only. The approach chosen for G.hn was to design a single transceiver optimized for multiple media, i.e. for power, phone and coax cables. Thus, G.hn transceivers are parameterized so that relevant parameters can be set depending on the wiring type. A parameterized approach allows to some extent optimization on a per media basis to address channel characteristics of different media without necessarily sacrificing modularity and flexibility. G.hn also defines an interoperable low complexity profile for those applications that do not require a full implementation of G.hn.

The G.hn WG engaged in a year long debate about the selection of the advanced coding scheme. The two competing proposals were based on a Quasi-Cyclic Low Density Parity Check (LDPC) code and a Duo-Binary Convolutional Turbo Code. The Turbo Code proposed in G.hn was meant to be an improvement over the one specified in the IEEE 1901 FFT-OFDM PHY/HomePlug AV as it allowed a higher level of parallelism and better coding gain. Following the comparative framework proposed in \cite{ITU:Jumbo:C-747, Galli2010fec}, the G.hn Working Group selected the LDPC code as the only mandatory FEC.

\subsection{PLC Coexistence} \label{sec.STDscoexistence}
PL cables connect LV transformers to a set of individual homes or set of multiple dwelling units without isolation. Signals generated within the premises interfere among each other, and with signals generated outside the premises. As the interference increases, both from indoors and outdoors sources, PLC stations will experience a decrease in data rate as packet collisions increase, or even complete service interruption. Hence, PL cables are a shared medium (like coax and wireless) and do not provide links dedicated exclusively to a particular subscriber. As a consequence, the PLC channel is interference limited and approaches based on Frequency Division Multiplexing (FDM) as in WiFi or coax are not suitable because only a relatively small band is available for PLC. As a consequence, it is necessary to devise mechanisms to limit the harmful interference caused by non-interoperable neighboring devices. Note that similar considerations can be made about the interference limited nature of many wireless networks, e.g. WiFi, WiMAX, Zigbee, Bluetooth, etc.

It is also important to ensure coexistence between Smart Grid and in-home BB technologies, since the former have traditionally a much longer obsolescence horizon than the latter. It is likely that the number of homes fitted with energy metering and control devices that utilize Smart Grid technology will dramatically increase in the near future. On the other hand, in-home BB technology continuously evolves, improving the transmission rate. The adoption of a coexistence mechanism will enable continued and efficient operation of Smart Grid devices in the presence of newly-deployed in-home BB devices.

The issue of PLC coexistence was first raised two decades ago in CENELEC. Since CENELEC does not mandate PHY/MAC recommendations, it was necessary to provide a fair channel access mechanism that avoided channel capture and collisions when non-interoperable devices operated on the same wires. In fact, if non-interoperable devices access the medium, then native CSMA and virtual carrier sensing do not work and a common medium access mechanism must be defined. CENELEC mandates a CSMA/CA mechanism only for the C-band \cite{STD:Cenelec1992} where a single frequency (132.5 kHz) is used to inform that the channel is in use. An extension of this method that utilizes three or four channel-in-use frequencies for HDR NB technologies operating in the the FCC band is now being discussed within PAP 15 \cite{Link:NIST-PAP-15}.

Another approach to coexistence was introduced by the HomePlug Powerline Alliance to solve the issue of non-interoperability between HomePlug 1.0 and HomePlug AV devices. The HomePlug \textit{hybrid delimiter} approach allows HomePlug AV/IEEE 1901 FFT-OFDM PHY stations to coexist with HomePlug 1.0 (TIA-1113) stations by pre-pending to their native frame the HomePlug 1.0/TIA-1113 delimiter. This allows stations to correctly implement CSMA/CA and virtual carrier sensing.

The hybrid delimiter approach is a CSMA-based coexistence mechanism and, thus, does not eliminate interference caused by non-interoperable stations and cannot guarantee QoS when the traffic of at least one of the coexisting technologies grows. Furthermore, the priority based QoS mechanism shared by HomePlug 1.0 and HomePlug AV/IEEE 1901 FFT-OFDM has been recently shown to be ineffective \cite{ChungJungLee2006HP1}. The use of hybrid delimiters is a somewhat inefficient approach if multiple technologies are to coexist as it would be necessary to pre-pend multiple delimiters (one for every non-interoperable technology) with increasing loss in efficiency. The HomePlug hybrid delimiter method also exhibits security weaknesses as it is not a mechanism based on \textit{fair sharing}. In fact, HomePlug AV/IEEE 1901 FFT-OFDM PHY can defer indefinitely HomePlug 1.0 (TIA-1113) stations from accessing the medium so that, while HomePlug 1.0 (TIA-1113) stations cease all transmissions, HomePlug AV/IEEE 1901 FFT-OFDM PHY stations remain the only active ones on the medium. This capability may raise security concerns since HomePlug AV/IEEE 1901 FFT-OFDM PHY stations (either legitimate or rogue) can stop from working Smart Grid devices based on HomePlug 1.0 (TIA-1113).

Except for the CSMA mechanisms described above, the issue of coexistence between BB-PLC devices has been rarely addressed in the technical literature and the first published paper dates back only few years \cite{TR:OPERA:Coexistence, ArlaBarbMata05, Stelts06, Dominguez06, GalKurOhu2009, LePhuLehn10cx}. The coexistence proposal by OPERA \cite{TR:OPERA:Coexistence} was meant to ensure compatibility between access and in-home BB-PLC deployments. The BB-PLC coexistence scheme developed in by the Consumer Electronics Powerline Alliance (CEPCA) and UPA \cite{Stelts06, Dominguez06} was to ensure coexistence between non-interoperable BB-PLC devices. In fact, since the lack of BB-PLC standards was causing a proliferation of proprietary solutions and since the industry did not seem to align behind any specific technology, coexistence seemed a necessary ``evil'' to ensure some etiquette on the shared medium and prevent interference. The CEPCA/UPA coexistence protocol is now included as an option in the IEEE 1901 Draft standard.

For the specific case of coexistence between the two IEEE 1901 PHYs, Panasonic proposed to the IEEE 1901 WG a novel coexistence mechanism called the Inter-PHY Protocol (IPP) \cite{GalKurOhu2009}. The IPP was initially designed to ensure compatibility with the CEPCA/UPA coexistence protocol but it was simpler, it allowed some distributed features, and also allowed devices to perform Time Slot Reuse\footnote{$^)$~Time Slot Reuse is the capability of nodes to detect when it is possible to transmit simultaneously to other nodes in neighboring systems, without causing harmful interference. Time Slot Reuse gains can be achieved also when the multiple stations sharing the medium are interoperable \cite{GalMasUra2009}.}. Although the IPP was originally designed to enable efficient resource sharing between devices equipped with either the IEEE 1901 Wavelet-OFDM or the IEEE 1901 FFT-OFDM PHYs, it was soon recognized that the IPP could have been also an excellent tool for regulating simultaneous access to the channel of both IEEE 1901 and non-IEEE 1901 devices, e.g. the ones based on the ITU-T G.hn standard. Panasonic modified the IPP originally conceived to extend coexistence to G.hn devices and proposed this enhanced mechanism called Inter-System protocol (ISP) to both ITU and IEEE. The ISP is now a mandatory part of the IEEE 1901 standard (see \cite{STD:IEEE1901:2010}, Chapter 16) and is also specified in ITU-T Recommendation G.9972 \cite{STD:ITU2010coexistence} which has been ratified by the ITU-T in June 2010. The approach followed in the design of the IPP/ISP is a radical conceptual departure from previous designs in CENELEC and in HomePlug which are both based on CSMA. Thus, none of the drawbacks mentioned above are present in the ISP  \cite{GalKurOhu2009}.

As a result of the efforts of PAP 15, IEEE 1901 compliant devices implementing either one of the two IEEE 1901 PHY/MACs can coexist with each other; likewise, ITU-T G.9960/9961 (G.hn) devices that implement ITU-T G.9972 can coexist with IEEE 1901 compliant devices implementing either one of the two IEEE 1901 PHY/MACs, and viceversa. A recent set of PAP 15 recommendations to the Smart Grid Interoperability Panel (SGIP) requires to \textit{``Mandate that all BB-PLC technologies operating over power lines include in their implementation''} coexistence. More in detail, in order to be compliant with this recommendation, IEEE 1901 compliant devices must implement and activate ISP (i.e., be always on), ITU-T G.9960/G.9961 compliant devices must be compliant with and activate ITU-T G.9972 (i.e., be always on), and any other BB-PLC technology must be compliant with and activate coexistence (i.e., be always on) as specified in ITU-T G.9972 or as in the ISP of IEEE 1901 \cite{TR:PAP15-BB-Recs}.

As there are already PLC technologies deployed in the field that do not implement ISP, it is important to understand to what extent the existing installed base of legacy technologies can create interference issues when ISP-enabled devices are deployed. The first consideration to make is that this is a minor issue as the installed based of BB-PLC devices is still very small when compared to other LAN technologies such as WiFi. Secondly, a lesser known but important benefit of ISP is its capability of eliminating in many cases of practical interest the interference created by an installed base of devices that does not use ISP but can be still controlled in an ISP-compliant manner under some mild assumptions \cite{ITU:CXlegancy:09GS-078}.

\section{The Role of Communications in the Smart Grid} \label{sec.RoleComms}
The history of communications through PLs shows that the power infrastructure is much more than the sum of its physical components. It is already a large scale cyber-physical system, where the physical system is coupled with a communication and computing network, in part aimed at controlling the automation aspects of the system, in part allowing the interaction and feedback of socio-economic networks through the energy market \cite{TR:NISTroadmap2009}. Initially, the electric system was composed of multiple but isolated generation plants. Recognizing that the interconnection between systems could provide higher profitability thanks to the access to a wider set of resources, the electric system was gradually transformed into an interconnected grid becoming the large scale cyber-physical system we know today. This transformation also introduced redundancy in case of equipment failure or unexpected demand fluctuations.

\subsection{Today's SCADA and Beyond} \label{sub.sec.SCADA}
The cyber infrastructure model that supports the management of the power network today is referred to as the Supervisory Control and Data Acquisition (SCADA) model. A system conforming to the SCADA model usually comprises the following components: a Human-Machine Interface, a supervisory SCADA Master server, a set of Remote Terminal Units (RTUs) and/or Programmable Logic Controller, sets of Intelligent Electronic Devices (IEDs), and the supportive communications infrastructure that furnishes the communications between the supervisory Master and the RTUs and between the RTUs and IEDs. The IEDs usually include various types of microprocessor-based controllers of power system equipment, such as circuit breakers, transformers, and capacitor banks. Multiple SCADA systems are today deployed within a plant and even at a substation. Thus, there is not a single SCADA network and some are based on Ethernet/IP and some are not.

The network support for SCADA has traditionally used combinations of wireless radio links, dial-up leased lines and direct serial or modem connections to meet communications requirements, although Ethernet and IP over SONET/SDH are more frequently used at supervisory control center or large substations. Although there is no single system, a two-level tree topology is very common to all communication networks supporting SCADA operations. Figure \ref{fig.SCADA} shows the RTUs at the intermediate level, sending control signals released by the supervisory master to the IEDs and sending measurement information from the IEDs to the supervisory master. Although newer substation automation systems are able to handle data generated at a faster pace, existing links between RTUs and the control center are often inadequate to handle an increasing volume of data  \cite{TomsBakkVenkBose2005}.

\insertonefig{8.8cm}{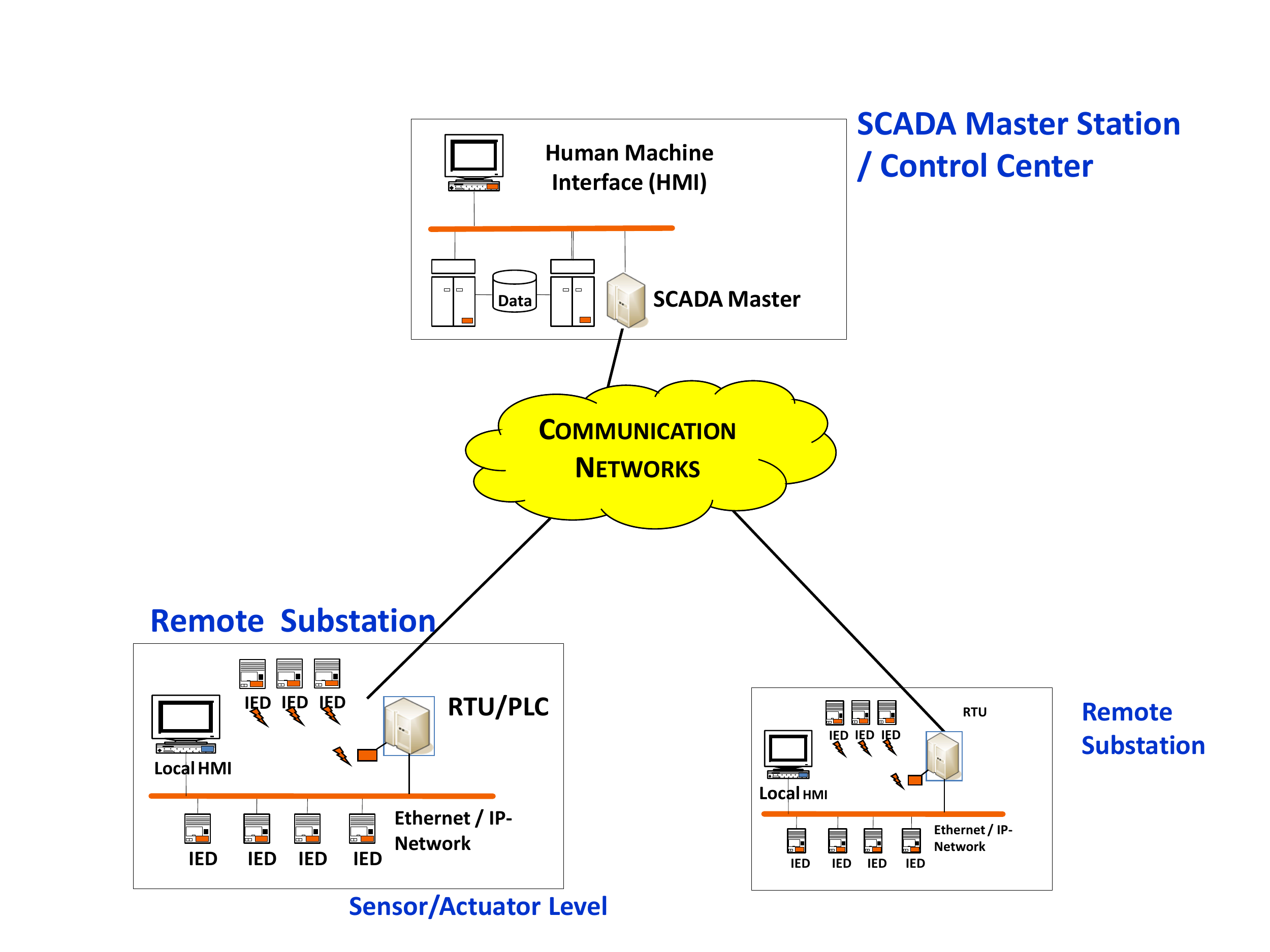}{\label{fig.SCADA}}{A Power Grid SCADA system.}

The SCADA centralized monitoring model is aimed at feeding data that constantly update the state estimation and system identification at the level of control stations, which assist the power system operator in his effort to adjust and/or optimize the power system operation and make sure that the system operational condition is a stable point for the system. The key problem of the SCADA model is, and has always been, the lack of architectural considerations on its latency, and what archetype for the information gathering would be needed to contain it.

Furthermore, in SCADA, most of the sensors capture and deliver measurements asynchronously. Hence, with SCADA the physical response of the system to contingencies cannot be optimally controlled in real time. In addition to the existing SCADA control, there are local feedback mechanisms in place such as generator excitation control, automatic governor control, automatic voltage regulator, HVDC-control, etc.

New transducers such as the synchrophasors or Phasor Measurement Units (PMUs) are being today deployed in the transmission side of the grid. PMUs can provide precise grid measurements of AC voltages and currents at high speed (typically 30 observations per second – compared to one every 4 seconds using conventional technology). Each measurement is time-stamped according to a common time reference, which utilizes the global positioning system (GPS) signal and has an accuracy better than 1 $\mu s$. Based on these measurements, improved state estimation can be derived so that it is possible to measure the state of a large interconnected power system. The immediate consequence of PMU deployment is that a large amount of data is being generated and the networking provisions for delivering this amount of data at the required QoS are not in place yet. The Wide Area Measurement System (WAMS) utilize a \emph{back-bone} phasor network which consists of phasor measurement units (PMUs) dispersed throughout the transmission system, Phasor Data Concentrators to collect the information and a SCADA master system at the central control facility. At the central control facility, system wide data on all generators and substations are collected regularly.

\subsubsection{Flexible AC Transmission System (FACTS)}
The FACTS is composed of power electronics and other equipment that provides control of one or more AC transmission system parameters to enhance controllability and increase power transfer capability of the network. FACTS control based on PMU can potentially be implemented as an effective wide area control means to mitigate sub-synchronous oscillations. This is a challenge to current SCADA/WAMS systems as measurements must be consistent and meet the real time requirement of fast transient and voltage stability control.

Distributed FACTS devices are smaller in size and less expensive in costs than traditional FACTS devices which may make them better candidates for wide scale deployment so that the topic of distributed control has been receiving increasing attention \cite{Divan_dFACTS:2007, Johal_dFACTS:2007, Rogers_Overbye_dFACTS:2008, Kechroud_Myrzik_dFACTS:2007}. Some researchers have also proposed that the control for Distributed FACTS devices could be decentralized as today more devices are equipped with fast communication capabilities and this scheme may help bypass the latency problem caused by the centralized monitoring and control implementation.

\subsubsection{Smart Grid in the Distribution Network}
Besides the increase in the data volume being generated within the transmission network for monitoring and control, there is another fundamental driver that will require a \textit{smarter} grid: the emergence of an increasingly dynamic and complex distribution side of the grid. The realization of an AMI, the integration of Renewable Energy Sources and other DER, and the new goals for improving distribution automation will produce radical changes in the distribution network. As a consequence, PMUs may soon find a role also in the state estimation of this new and dynamic distribution grid because of the higher level of uncertainty due to the integration of time-varying DER and distributed control mechanisms.

The creation of a pervasive AMI has polarized considerable attention as many advocate the AMI network as being the core sensing and measurement system of the distribution network.
Most proposed AMI architectures include a data hub or concentrator service where measurement data from smart meters will first be collected and unified before being further sent to the back office of the utility. Since this centralized model does not scale, it is reasonable to look at alternative architectures that have greater parallelism in designing the next generation cyber system. The availability of scalable networking alternatives as well as decentralized and fully automated processing will allow connecting the embedded intelligence in the system in a way that will support each of the physical devices with real time feedback from its neighboring devices. This is a profound paradigm shift from current remedial practices that merely change generator settings thus effecting everything downstream rather than a particular portion of the grid that has a problem. Going forward it will be necessary to understand Volt/VAr not only at a macro level but also at micro level over each segment of the grid.

\subsection{Control and Sensing for Cyber-Physical Systems} \label{sec.ControlSensorNetworking}

The grid is a complex cyber-physical infrastructure composed of a maze of interdependent and interacting parts. The cyber-physical system can be seen as pair of partner networks: a physical network over which energy flows and the cyber system (including a wide area network of sensors and data sinks that compute and relay information to actuation sites). The service provided is energy delivery from sources to destination and, in principle, nothing else but the physical network is required. If a physical part of the system fails, the safe operational limits for the network may change. Hence, the timely notification of failures is critical and, as failures spread, the network will face a sudden surge in highly correlated sensor traffic similar to a broadcast storm \cite{broadcast-storm}. Effectively the sensors are reporting the same event, however in doing so they will compete for network resources, causing congestion.

The SCADA model leaves a great deal of control to the human operator in the loop. Can one develop a fully automated solution?
As Witsenhausen's counterexample indicated, the separation of estimation and controller design fails to
hold even in the simplest settings \cite{witsenhausen}. Many theories are emerging that deal with the issue of control under communication constraints that apply broadly to cyber-physical systems (see e.g. \cite{journal:mitter_2001, conf:elia_2004, tatikonda, journal:sinopoli_2006, journal:haspanha_2005}), however modular and scalable solutions of network control are still elusive in many cases. These technical obstacles are especially relevant when a separation of time scales is impossible.  Unlike transportation, water network and other commodities that are encountered in large scale supply chains, electrical power moves just as fast as communication signals do. Therefore, both the physical network dynamics and the cyber system data spread at comparable speeds, exacerbating the difficulties of decoupling communications from control and management.

Part of the the difficulty in the optimization of concurrent controllers is that each controller can infer information about unobservable events not only by pooling sensors information but also by observing the other controllers actions \cite{ricker2}. However, in some important cases the controllability of a discrete event system is undecidable \cite{Tripakis}.

Recently, the low cost of communication and computation devices has determined a considerable pressure to grow these networks in size and complexity. Sensor networking research has flourished in the past ten years \cite{Estrin} and a new generation of sensor communications is being standardized at a fast pace especially in the wireless field - e.g., see IETF 6LoWPAN, IEEE 802.15.4, IEEE 1451, etc. PLC standards are emerging with similar functionalities, as discussed in Section \ref{sec.PLCforSmartgrid}. Remarkably, this process has reversed completely the natural order of design, where the information network infrastructures, routing and clustering primitives are chosen from rather generic sensor networking models that are not delay sensitive nor specifically tied to energy distribution specifically, thus dictating ultimately the delays that the control needs to work with, not the other-way around.

It is therefore likely that this first generation of devices, networks, data processing and software agents will be over-designed in many ways and also lacking in other aspects that are today unforeseen. This will create incentives in designing a new generation of optimized devices and protocols tailored to the actual Smart Grid needs. Some important elements that these new solutions will have to incorporate are considered in the following subsections.

\subsubsection{Grid Control Aspects}
The voltage on the mains is a narrow bandpass signal, around the mains frequency $f_0=$ 60 or 50 Hz. The complex phasor vectors $V$ and $I$
\begin{equation}\label{eq:phasorVI}
 \begin{array}{cc}
   V = \textrm{V}\angle V \\
   I = \textrm{I }\angle I,
 \end{array}
\end{equation}
represent the sinusoids of instantaneous vector voltage $v(t)$ and injected vector current $i(t)$ respectively, around $f_0$:
\begin{equation}\label{eq:instantaneous_vi}
\begin{array}{cc}
v(t) = \sqrt 2 \textrm{V} \cos(2\pi f_0 t + \angle V), \\
i(t) = \sqrt2 \textrm{ I} \cos(2\pi f_0 t + \angle I).
\end{array}
\end{equation}
In the NB regime the power network dynamics are coupled by the algebraic equation
\begin{equation}\label{eq:YU=I}
YV=I,
\end{equation}
where  $Y$ is the matrix of network admittances at frequency equal to $f_0$, which is determined not only by the connecting topology but also its electrical parameters. The relationship in (\ref{eq:YU=I}) is valid because the variations of $Y(f)$ over the spectrum of $v(t)$ are negligible.  Given a network with $n$ nodes and $m$ links (which may also be referred to as ``buses and branches (or lines)'' in power grid analysis; or ``vertices and edges'' in graph theory and network analysis), each link $l = (i,k)$ between nodes $i$ and $k$ has a line impedance at 60 Hz $z_{pr}(l)=r(l)+jx(l)$, where $r(l)$ is the resistance and $x(l)$ the reactance. Usually, for HV transmission network, the reactance dominates. The $n \times n$ network admittance matrix $Y$ is
\begin{equation} \label{eq:Y1}
   Y=A^Tdiag(\mathbf{ y_{pr}})A
\end{equation}
where $\mathbf{y_{pr}}$ is the line admittance vector, whose elements are $y_{pr}(l)=1/z_{pr}(l)$, and $A$ is
the line-node incidence matrix. Each bus corresponds to a certain power flow injected (generator bus) or absorbed (load bus), or it simply represents an intermediate bus. The instantaneous power at each bus is given by $p(t) = v(t)\odot i(t)$, where $\odot$ means vector element-wise multiplication. Taking the phasor value, we have
the network power flow equation as
\begin{equation} \label{eq:S}
S = V \odot I^*,
\end{equation}
where $(\cdot)^*$ indicates complex conjugation, and $S = P+jQ $ is the vector of injected complex power
\begin{equation} \label{eq:PQ}
 \begin{array}{cc}
  P=\textrm{Re}(V \odot I^*) = \textrm{V}\textrm{I} \cos (\angle V-\angle I) \\
  Q=\textrm{Im}(V \odot I^*) = \textrm{V}\textrm{I} \sin (\angle V-\angle I).
 \end{array}
\end{equation}
where $P$ is the \emph{real power} or \emph{active power}, which is equal to the DC component of the instantaneous power $p(t)$; whereas $Q$ is the \emph{reactive power} which corresponds to the $2f_0$ sinusoid component in $p(t)$ with zero average and magnitude $Q$.

A set of basic constraints needs to be satisfied for enforcing stability in the power grid: (a) the network power flow must be balanced; (b) the input power for generation or loads adjustment or power injects from other kinds of sources must have strict operational ranges; (c) voltage must take acceptable levels; (d) line thermal limits must be enforced, i.e. line current should keep its magnitude below a specified limit; (e) stability condition must be satisfied, i.e., the Jacobian matrix $J(V)$ of the network power flow equations must have negative real parts which keep a safe distance from zero.

Mathematically, the conditions described above can be written as follows\footnote{$^)$~Note that $S$, $V$, and $Y$ are time-varying variables, since their magnitude and/or phase angles are changing with system operating status. Here we omit the ``$(t)$'' term only for notation conciseness.}:
\begin{eqnarray}
  \nonumber (a)&& S = V \odot (YV)^* \\
  \nonumber(b)&& S_{\min} \le S \le S_{\max} \\
  (c)&& \textrm{V}_{\min} \le \|V\| \le \textrm{V}_{\max} \\
  \nonumber (d)&& \|diag(\mathbf{ y_{pr}})A V \| \le \textrm{I}^{line}_{\max} \\
  \nonumber (e)&& \textrm{Re} \left( eig(J(V)) \right) \le -\varepsilon
\end{eqnarray}

\noi where the Jacobian matrix $J(V)$ of the network power flow equations is defined as follows:

\begin{equation} \label{eq:Jacobian}
J(V)=\left[
 \begin{array}{cc}
 \frac{\partial P}{\partial \angle V} & \frac{\partial P}{\partial \|V\|} \\
 \frac{\partial Q}{\partial \angle V} & \frac{\partial Q}{\partial \|V\|}
 \end{array}
\right].
\end{equation}

The key feedback mechanisms consist in controlling the elements of $S$, by increasing supply or shedding loads, or controlling $Y$ by switching parts of the infrastructure or utilizing FACTS. In this case the monitoring needs to be synchronous; polling each part of the network and gathering centrally all the data, and then distributing the control signal is a solution that is not scalable and may result in congestion.

\subsubsection{Traffic Generated by the Physical System}
There is a universal brute force solution to congestion problems: increase the service rate so intermediate nodes buffers never grow. In so doing, network bottlenecks will not constitute a problem since over-provisioned nodes will push through the messages received in face of the worst conditions. There is clearly merit in this view of the problem, as technology that offers high rates becomes cheaper and one does not need to explore new networking concepts to design Smart Grid. The approach of over-provisioning would certainly help infrastructure monitoring and the wide area control, but would also entail the higher cost of provisioning higher than necessary capacity links.

An alternative view that has emerged in the sensor networking community is to exploit directly the data structure and correlation among the sensor data to reduce the information flows and to manage the rise in complexity of routing and processing data \cite{scaglione_servetto_2003, chou03, duarte-melo03, cristescu, sung, pattern08}. In fact, eliminating queuing delays only may be insufficient since the brute force solution of polling all the sensors can become the true bottleneck for the sheer problem of collecting all the data in a timely way from the sensors.

Modeling the traffic of phase sensors in the electrical network is an important research direction. Network scientific work on the power network infrastructure has so far been focused on capturing topological characteristics and studying vulnerability to topological changes of the transmission grid. The models by Watts and Strogatz (1998) \cite{watts98}, Newman (2003) \cite{newman03}, Whitney and Alderson (2006), fall in this class. This approach is important when dealing with PLC for Smart Grid and utility applications because the communication graph is a subgraph of the physical infrastructure used for power delivery. The same network scientific analysis carried out to analyze the transmission grid can then be utilized also to provide insights on the network coverage of a PLC based Smart Grid distribution system. Building on this work, we present in Sect. \ref{sec.Zhifang} an analysis that captures accurately both topological and electrical characteristics of the distribution power grid. This approach allows the study of the optimal placement of PMUs to allow continuous monitoring of channel and network states and provide information on the physical state of the distribution system.

\subsubsection{Cooperative Schemes for PLC Networks}
Naive solutions based on polling (see IEEE 1901 Draft, Chapter 8) simply do not scale.
Given its wide geographical area of deployment, the Smart Grid will utilize relays potentially at a massive scale. It is well known that interference can lead to vanishing throughput as the size of the network scales up, as shown in \cite{GuptaKumar2000} for the case of wireless networks. PLC as well is likely to be equally challenged due to the fact that relays interfere with each other. This complicates greatly routing decisions. In general, routing itself will also need to be flexible; it is, in fact, critical to equip the network with scalable primitives for self-organization that would allow the network to find rapidly alternative paths to deliver sensitive information, in light of local failures.

When using relays, a problematic operation in broadcast media is, paradoxically, broadcasting (or multicasting). In order to use decentralized storage, control microgrids and taper off the demand as a means to compensate for volatility of supply, broadcast control signals will very often flow through the network. Delivering in a timely fashion these messages to a large population of Smart Grid terminals through many relays will produce a broadcast storm if protocols to support this function are not designed judiciously \cite{broadcast-storm}. Failures in the infrastructure are likely to generate a similar storm of signals, due to cascading effects that impact close by elements of the system. The classical solution to this problem at the network layer is either forming a static routing table that resolves such conflicts (this takes time and is not robust), or resorting to the so-called probabilistic routing \cite{broadcast-storm}. Interestingly, these functions can be greatly improved upon by using physical layer cooperation in forwarding the signal. Cooperation is a physical layer solution to the relay problem, it allows signals to be superimposed in the time and frequency dimensions by appropriately encoding as well as timing the signals transmitted by populations of relays. This concept has been independently introduced for wireless networks and for PLC networks in \cite{ola} and \cite{bumiller:2002}, respectively. The first working implementation of a cooperative PLC-based AMI system using HDR NB-PLC was realized under the European REMPLI project (Real-time Energy Management via Power lines and Internet) \cite{Link:REMPLI, BumiLampHras2010rempli}. The REMPLI project has experimentally demonstrated the possibility of using HDR NB-PLC in transforming channel contention into channel \textit{cooperation} by using a Single Frequency Network with flooding based routing. The advantage of these approaches is that the delivery of the message can be predicted much more accurately and the transmission is more power efficient.

\section{The Role of PLC in the Smart Grid}\label{sec.PLCforSmartgrid}
There are many examples of applications where PLC can be used for utility applications. In the next subsections, we will review the salient applications of PLC for the Smart Grid at all voltage levels - from HV lines down to and within the home.

\subsection{PLC for High Voltage Networks}
Although the greatest transformation from today's grid to tomorrow's Smart Grid is expected to take place mostly on the distribution side, also the transmission side will have to undergo progressive changes which some believe will be slower than for the distribution side and will also occur at an evolutionary pace \cite{HoroPhadRenz2010}. The availability of a reliable communication network on the transmission side is critical for the support of several applications such as state estimation (PMU over WAMS), protective relaying, SCADA expansion to remote stations, remote station surveillance, and power system control. There are established PLC technologies operating over AC and DC HV lines up to 1,100 kV in the 40-500 kHz band that allow data rates of few hundred kpbs and play an important role in HV networks due to their high reliability, relatively low cost and long distance reach \cite{TR:ABB:PLCsingHV}. HV lines are good waveguides as channel attenuation characteristics show a benign pass-band and time-invariant behavior - see Table \ref{tab.PathLoss}. The noise is mainly caused by corona effect and other leakage or discharge events, and corona noise power fluctuations of some tens of dB can be observed due to climatic dependency. Compared to LV/MV, HV lines are a much better communications medium characterized by low attenuation - see Table \ref{tab.PathLoss}. When available, alternative communications technologies used on the transmission side of the grid are based on either fiber optical or microwave links which also allow higher data rates than PLC.

The first PLC links over HV lines were installed in the early 1920s with the goal of providing operational telephone services and were based on analog Single Side Band Amplitude Modulation \cite{Schwartz2008historyPLC}. The first digital PLC system was introduced by ABB in 1999 allowing data rates up to 64 kbps in a 8 kHz band \cite{TR:ABB:PLCsingHV}. Today, the state of the art of HV digital modems support data rates of 320 kbps in a 32 kHz band and a reach of 100 km. Note that this is a very high spectral efficiency (10 bits/s/Hz), which is 50\% higher of what BB-PLC can achieve ($<7$ bits/s/Hz) or nearly an order of magnitude more of what NB-PLC are capable of (\texttildelow 1 bit/s/Hz). Today, the use of PLC over HV lines is well established and thousands of links have been installed in more than 120 countries for a total length of some millions of kilometers. Digital PLC over HV lines has not been standardized yet, but a couple of years ago the IEC TC57/WG20 started to work on updating the obsolete analog PLC standard IEC 60495 to include digital PLC for HV.

Besides providing connectivity on the transmission side, PLC over HV lines can also used for remote fault detection. For example, successful experiments were recently reported for the detection of broken insulator, insulator short circuit, cable rupture, and circuit breaker opening and closing \cite{AdamSilvMart2009hv}. In another example, PLC over HV appears to be also useful in determining the change in the average height above ground of horizontal HV overhead conductors. Authors in \cite{VillCloeWede2008hhv} report successful testing on a 400 kV overhead HV line of a real-time sag monitoring system based on PLC in the FCC band.

There is today a growing interest in achieving higher data rates via PLC over HV lines. The feasibility of sending high data rate PLC signals over HV lines has been reported recently by the US Department of Energy, American Electric Power, and Amperion, who jointly tested successfully a BB-PLC link over a 69 kV and 8 km long line with no repeaters \cite{TR:DoE:AmperionHV}. Data rates of 10 Mbps with latency of about 5 ms were reported while complying with FCC emission limits. The trial employs multiple 5 MHz bands in the range of 2-30 MHz using DS2 (UPA) chips to communicate over two contiguous 69 KV lines \cite{PrivComm:AmperionNov2010}. The trial has also successfully tested an important application of BB-PLC over HV: protective relaying using Current Differential Protection \cite{PrivComm:AmperionNov2010}. Current Differential Protection systems have been traditionally supported using fiber-optic links so that the successful use of BB-PLC is an important result that could help greatly reduce the cost of a vital line protection technology. Next steps for this project is to raise the applicable voltage to 138 kV and also extend repeater spacing. Other  international activities involving PLC over HV lines can be found in \cite{PighRahe05hv, HyunLee08hv, AquiGutiPijo2009hv}.

At this time, it is possible to express only cautious optimism about the use of BB-PLC in the transmission side as further testing and validation is needed to bring BB-PLC over HV to a commercial stage.

\subsection{PLC for Medium Voltage Networks}
An important requirement for future Smart Grids is the capability of transferring data concerning the status of the MV grid where information about state of equipment and power flow conditions must be transferred between substations within the grid. Traditionally, substations at the MV level are not equipped with communication capabilities so the use of the existing PL infrastructure represents an appealing alternative to the installation of new communication links. Some substation automation functions need substation IEDs to communicate with one or more external IEDs. In the case of fault location, fault isolation and service restoration, substation IEDs must communicate with external IEDs such as switches, reclosers, or sectionalizers. In another example, implementation of voltage dispatch on the distribution system requires communication between substation IEDs and distribution feeder IEDs served by the substation. All these communications require low-speed connectivity that is well within PLC capabilities.

A large portion of MV equipment in the world has been installed more than 40 years ago. Fault detection as well as monitoring for ensuring longer lifespan to critical cable connections is then becoming a true operational, safety and economical necessity. Most techniques used today include on-site expensive truck rolls; for example, available power cable diagnostics are based today on partial discharge measurements (typically based on Time Domain Reflectometry) on temporally disconnected connections which are externally energized. From an operational point of view online diagnostic tools are preferable and soon will become the main trend \cite{Mich03}. The coupling of PLC signals up to 95 kHz (European CENELEC A-band) for online diagnostic data transfer over MV cables is studied in \cite{WoutWielVeen05} where the authors also emphasize the advantage of integrating diagnostics tools that serve the dual purpose of sensing and communication devices.

DG systems can supply unintentional system islands isolated from the remainder of the network. It is important to quickly detect these events, but passive protections based on traditional measures may fail in island detection under particular system-operating condition. The use of LDR NB-PLC (CENELEC A-band) for injecting a signal in the MV system has been analyzed and tested in \cite{BenaCaldCese03}, and it appears to be less expensive compared to other methods based on telephone cable signals. A similar approach has been investigated in \cite{RoppAakeHaig00} for the prevention of islanding in grid-connected photovoltaic systems and it was found that PLC-based ``\textit{islanding prevention offers superior islanding prevention over any other existing method}.'' Other applications of PLC within the area of DG can also be found in \cite{BenaCald07}.

In addition to remote control for the prevention of the islanding phenomenon, other applications related to monitoring on the MV side (temperature measurement of oil transformers, voltage measurement on the secondary winding of HV/MV transformers, fault surveys, power quality measurement) have also been discussed and analyzed \cite{CataDaidTine08}.

\subsection{PLC for Low Voltage Networks}
Most PLC Smart Grid applications on the LV side are in the area of AMR/AMI, vehicle-to-grid communications, DSM, and in-home energy management. Those applications will be addressed in the next sub-sections.

\subsubsection{Automatic Meter Reading and Advanced Metering Infrastructure} \label{sec.AMR-AMI}
In addition to basic one-way meter reading (AMR), AMI systems provide two-way communications that can be used to exchange information with customer devices and systems. Furthermore, AMI enables utilities to interact with meters and allows customer awareness of electricity pricing on a real-time basis \cite{TR:NISTroadmap2009}. Although smart meter deployment is getting today a lot of attention worldwide, a smart meter is not really a necessary part of the Smart Grid as there are several alternative ways to implement Smart Grid applications without smart meters. On the other hand, smart meters are important tools for the utilities to reduce their operational costs and losses because they provide capabilities that go beyond simple AMR, such as remote connect/disconnet and reduction of the so called non-technical losses, e.g. losses due to energy theft.

PLC technology is certainly well suited for AMR/AMI. There is a vast amount of field data about the performance of PLC-based smart meters as few hundred million UNB/NB-PLC devices have been deployed around the world.

As mentioned in Sect. \ref{sec.UNBandNB}, UNB-PLC devices were the first ones to be used for AMR/AMI. Although UNB-PLC systems are characterized by very low data rates, UNB-PLC signals propagate easily through several MV and LV transformers. Furthermore, UNB-PLC does not require any kind of PL conditioning as other PLC technologies operating at higher frequency would often require due to the low pass effect of shunt power factor correction capacitors and series impedances of distribution transformers. As a consequence, these systems are able to cover very large distances (150 km or more). In the last couple of decades, UNB-PLC system have experienced good success in the market. The Turtle System has found good applicability in those areas served by US rural cooperatives as they are characterized by low population density and wide geographical spread. Several million TWACS-based end-points have been deployed in rural as well as in urban and sub-urban areas located in the US and Latin America and provide meter reading at 15 minute intervals \cite{PrivComm:TWACSnov2010}.

Also NB-PLC technologies are gaining interest for AMI applications, an interest exemplified by the recent creation of two projects devoted to the standardization of HDR NB-PLC transceivers (IEEE 1901.2 and ITU-T G.hnem). The capability of HDR NB-PLC of delivering substantially higher data rates with respect to UNB-PLC comes at the price of reduced range and, sometimes, transformer conditioning. Not all PLC technologies offer the same reliability in passing the distribution transformer and often this capability strongly depends on the transformer itself. The equivalent circuit of a transformer contains both capacitors and inductances, where the capacitors appear as shunts and this produces the well known low-pass behavior. However, the combination of various capacitors and inductances should give rise to a resonant behavior at various frequencies thus adding frequency selectivity on top of the low pass trend. This should be a general behavior at all frequencies, although at high frequencies other effects such as RF coupling may also appear. This behavior was recently confirmed in \cite{Black10transf} where it was reported that transformers offer several narrowband windows of low attenuation from the Low Frequency up to the Very High Frequency region. Thus, even though BB-PLC signals do not pass through (or around) the distribution transformer and broadband connectivity between MV and LV necessarily requires the installation of coupling units to by-pass the transformer, low data rate communication between two BB-PLC nodes located on the two sides of a distribution transformer may be sometimes possible without by-pass couplers. These windows of low attenuation are present also at lower frequencies thus allowing NB-PLC technologies to pass the transformer in some cases - most likely when there is high frequency diversity like in HDR NB-PLC so that multiple windows of low attenuation fall in the communication bandwidth \cite{RazaUnarKama2010g3}. These characteristics call for sub-banding techniques and frequency agility capabilities in PLC transceivers. Although these results are encouraging, it is difficult to draw at this time general conclusions on this matter since there is no statistical model for transformers that allows a more quantitative assessment of the capability of PLC signals to pass through (or around) the distribution transformer.

The architectural consequence of MV/LV connectivity is that many more meters could be handled by a single concentrator located on the MV side. This concentrator node would then send the aggregated data from many meters back to the utility using either PLC or any other networking technology available in situ. This capability also heavily impacts the business case when there is a very different number of customers per MV/LV transformer: in North America, the majority of transformers serves less than 10 customers; in Europe, the majority of transformers serves 200 customers or more. Thus, especially in the US, it is economically advantageous to avoid coupler installation and resort to technologies that allow connectivity between the MV and LV sides - and possibly also between meters served by different distribution transformers (LV/MV/LV links). When there are very few end-points (meters) per distribution transformer as in the US, it is convenient to push the concentrator up along the MV side (and even up to the substation) and handle multiple LV sections so that more end-points can be handled per concentrator. The low number of end-points per transformer in the US makes UNB-PLC solutions like TWACS attractive as the concentrator is located in the substation and can handle a large number of meters with no additional communication infrastructure, e.g. repeaters or couplers, between substation and meters. On the other hand, the large number of end-points per transformers in Europe does not really require to locate the concentrator up in the substation or on the MV side as it can be conveniently located on the LV section of the grid. Thus, the capability of a PLC technology to pass through distribution transformers may be more appealing in the US rather than in Europe.

In emergency situations it is often the case that conventional networking technologies encounter congestion due to a spike in the collision rate, i.e. when all meters tend to access the channel at the same time (blackout, restoration, etc.) or when multiple DR signals requiring immediate action are sent to households. In these challenging scenarios, traditional networking approaches including wireless sensor networks fail due to the network congestion and competitive channel access mechanism.
Unlike wireless solutions based on ZigBee or WiFi, PLC-based AMI have a proven track record of being able to avoid network congestion when cooperative schemes are employed - see the REMPLI project \cite{Link:REMPLI}, \cite{BumiLampHras2010rempli}.

\subsubsection{Vehicle-to-Grid Communications}
A PHEV charges its battery when connected to an Electric Vehicle Supply Equipment (EVSE) which, in turn, is connected to premises wiring or to distribution cables (airport, parking lots, etc.). A variety of applications scenarios can be envisioned in enabling a communication link between the PHEV and the utility, e.g. for the control of the localized peak load that the increasing penetration of PHEVs would inevitably create. The availability of a communication link between the car and the EVSE (and even beyond the EVSE to the meter, the Internet, the HAN, the appliances, the utility, etc.) will be the key enabler for these applications.

The first distinctive advantage of PLC for vehicle-to-grid communications is the fact that an unambiguous physical association between the vehicle and a specific EVSE can be established, and this is something that is not possible to accomplish with wireless solution even if short range. This physical association has advantages, especially in terms of security and authentication. Although the PLC channel in this scenario is impaired by several harmonics present due to the inverter, there are today several ongoing tests on both BB-PLC and NB-PLC solutions within the ``PLC Competition'' being conducted by the Society of Automotive Engineers (SAE). In terms of cost, worldwide regulations, and ease of upgrade, NB-PLC solutions offer some advantages with respect to BB-PLC as argued in detail in Sect. \ref{subsec.BestPLC}. Since NB-PLC are also excellent choices for meters and appliances, the availability of a single class of PLC technologies for the inter-networking of different actors is of course tempting.

\subsubsection{Demand Side Management (DSM)}
One of the primary DSM applications on the LV side is DR which has been receiving growing interest, especially in the US \cite{HoppGoldBhar2007SummerDR, FERC:DemandResponse:2009}. DR refers to the ability to make demand able to respond to the varying supply of generation that cannot be scheduled deterministically, e.g. solar and wind. Thus, DR is a means to alleviate peak demand and to bring more awareness on energy usage to the consumer \cite{TR:NISTroadmap2009}. It is believed that DR will allow a better control of peak power conditions, maximize the use of available power, increase power system efficiency through dynamic pricing models, and allow customers to participate more actively to energy efficiency. Implementation of DR requires establishing a link (either direct or indirect, e.g. via gateway) between the utility and household appliances.

The largest direct load control system in the world has been operating in Florida for over twenty years using a UNB-PLC technology (TWACS). Florida Power and Light manages via TWACS over 800,000 Load Control Transponders installed at the premises of over 700,000 customers and can shed up to 2 GW of load in a matter of a few minutes. Florida Power and Light has also deployed 1.4 million TWACS-enabled end-points for AMI \cite{PrivComm:TWACSnov2010}. It is interesting to verify that such large scale DR/AMI systems can operate successfully using a communications system characterized by very low data rate.

Due to the higher attenuation that PLC signals experience over the LV side, BB-PLC solutions may not always be ideal for DR applications when direct load control is implemented since the distance between appliances and the utility signal injection point (the smart meter, the MV/LV transformer) may be in some cases too large. On the other hand, when DR is implemented with indirect control via a gateway, e.g. a Home/Building Energy Management System (HEMS/BEMS), then BB-PLC solutions are technically adequate and would provide the added benefit of being able to transfer securely data from Smart Grid applications to the HAN and vice versa. Although technically adequate, other considerations related to cost may arise as BB-PLC technologies may be overly-dimensioned for carrying out DR. Due to the much lower path loss at lower frequencies, NB-PLC solutions are also good candidates for DR applications for both direct and indirect load control.

\subsubsection{In-Home Environment}
There are intriguing possibilities of tying Smart Grid applications with HEMS, and there is a strong belief that these application will help foster a behavioral change in how consumers address energy consumption. The home is a natural multi-protocol and multi-vendor environment and it is unrealistic that this will change anytime soon even though there is a lot of pressure by some industry segments to reduce the number of allowed networking choices. A variety of BB-PLC solutions will continue to be installed by consumers regardless of any convergence in the networking choices for the Smart Grid. From this point of view, segregating Smart Grid applications in one band (CENELEC/FCC/ARIB) and separating them from traditional entertainment and Internet access ones running on BB-PLC (but also with the capability of linking these applications securely via the HEMS) seems a good engineering solution that balances efficiently the various requirements of these very different applications. However, the use of NB-PLC in the in-home environment may require special attention to cope with reduced cross-phase connectivity since the capacitive nature of cross-phase coupling yields higher attenuation at lower frequencies than at the higher ones used in BB-PLC \cite{SugiYamaKata08nb}.

Although there is evidence that a HEMS does not provide compelling financial benefits to residential customers, there is substantial evidence that a HEMS can yield substantial benefits to utilities in terms of improving grid reliability as well as reducing peak demand. In fact, a HEMS can serve the function of ``sensor'' in a much more complete and effective way than what a smart meters would be capable of doing. While the smart meter is a low-cost sensor and can only report instantaneous demand, a HEMS could actually report to the utility (or third party energy service provider) the forecasted demand of energy and provide more complex sensing functions. For example, the forecasting capability of a HEMS could be very accurate as it would be based on the ``state'' of the home and on the behaviorial model built on consumer activity. The state of the home tracked by a HEMS could include: the present and predicted energy demand of an appliance as it goes though its service cycle, storage levels of batteries, amount of consumer shifted demand (service queue), etc. If a utility had at its disposal the knowledge of the state of every home (or of a set of homes or microgrids via aggregators), forecasting and scheduling of generation and DSM would be possible with more relaxed communications requirements. Furthermore, storage levels and queued demand could also become part of pricing models \cite{RoozDahlMitt10sgcom}.

We also point out that today there is a growing interest in hybrid AC/DC wiring infrastructure. Within the home, the development of a DC infrastructure yields great benefits to energy generation (photovoltaic, fuel cell) and storage (rechargeable battery). Both NB and BB-PLC greatly benefit from operating over DC lines as the channel is time-invariant and appliance cyclostationary noise disappears - with the exception of impulsive noise caused but AC/DC inverters.

\subsection{What PLC Technology Fits Best Smart Grid Applications?} \label{subsec.BestPLC}
UNB, NB and BB-PLC solutions can find their space of application and the choice of which PLC technology best fits the application scenario will depend not only on technical matters but also on regulatory and business case aspects. In fact, regulations on allowed emissions levels and available frequencies can make us reach different conclusions on what PLC technology is preferable for a given scenario \cite{GebWeiDos2003}. For example, FCC Part 15 in the US allows the use of both NB and BB-PLC technologies in outdoor deployments; in the EU, on the other hand, BB-PLC solutions may not be practical because of stricter regulations that limit the allowable transmit power and, as a consequence, would require smaller repeater spacing and thus increased deployment costs. In another example, the use of BB-PLC solutions outdoor is also forbidden in some countries, e.g. Japan where only UNB or NB-PLC solutions would be available for Smart Grid applications.

One compelling advantage of using PLC is that the traditionally separated functions of sensing and communicating blur together and thus a PLC transceiver could be designed to switch between functioning as a sensor and as a modem. This capability may have applications in Power Quality which is an important concern for utilities because of the value of predicting and avoiding electric disturbances \cite{RibeSzczIrav07}. Beyond Power Quality, PLC can be used to reveal unhealthy grid devices (e.g.cracked insulators, broken strands, etc.) that emit noise (often in the High Frequency band) prior to eventual failure. Another advantage of using existing PLs as a communication channel is that utility applications almost always require redundancy in protection and control applications, and the need for redundancy should also be extended to the availability of redundant communication channels \cite{KaszHuntVazi07}. From this point of view, the availability of an existing wired infrastructure greatly reduces the cost of deploying a redundant communication channel. An additional advantage in the use of PLC for Smart Grid applications is that PLs often represent the most direct route between controllers and IEDs when compared to packed switched public networks - see also the considerations made in Sect. \ref{sec.Zhifang} on the topological characteristics of the transmission side of the power grid. This is advantageous when dealing with applications such as tele-protection since ensuring a bounded low latency is very important. Last but not least, PLs provide a communication path that is under the \textit{direct and complete} control of the utility which is an important aspect when a utility operates in a country with de-regulated telecommunication markets.

\begin{table}[!h]
  \centering
  \caption{Typical path loss values for PLC in dB/km. Values may vary depending on cable type, loading conditions, weather, etc. OH: overhead; UG: underground.}
  \label{tab.PathLoss}
\begin{tabular}{lccc}
  \hline\hline
                        & $f=100$ kHz & $f=10$ MHz \\ [0.5ex]
\hline
  Low Voltage           & 1.5-3         & 160-200 \\
  Medium Voltage (OH)   & 0.5-1         & 30-50  \\
  Medium Voltage (UG)   & 1-2           & 50-80  \\
  High Voltage (OH)     & 0.01-0.09     & 2-4  \\ [1ex]
  \hline \hline
\end{tabular}
\end{table}

Of course, the cost savings of having the infrastructure available should be weighed against the cost of deployment of repeaters and couplers. Though it is hard to give universal values for path loss since many factors influence it (overhead or underground cables, type of cables, loading, weather, etc.), typical values of path loss for the PLC channel in dB/km are given in Table \ref{tab.PathLoss} - see for example \cite{Book:Dostert2001, TR:OPERA:Pathloss, Olsen05, STD:IEEE643-2004:2010, PrivComm:AmperionNov2010}. Additional data collected in the US suggest that the average in-home channel attenuation encountered by BB-PLC transceivers ranges between 40 and 50 dB for urban and sub-urban homes, respectively \cite{Link:Galli2011wirelinemodel}. As Table \ref{tab.PathLoss} suggests, the use of BB-PLC over LV networks can entail very small repeater spacing due to the high path loss whereas larger repeater spacing can be tolerated over HV/MV networks, especially for the overhead case. Due to the wide variability of scenarios, PLC may be a good solution or not and its appropriateness must be assessed on a case-by-case basis - just as one would do for any other communication technology.

NB-PLC has several advantages when compared to BB-PLC when AMR/AMI or DR applications involving appliance control are considered - and even when NB-PLC solutions are compared with scaled down versions (low complexity, low power, low data rate) of BB-PLC solutions\footnote{$^)$~An example of scaled down version of BB-PLC devices is the low complexity profile defined in the ITU-T G.hn standard. A non SDO-based example is given by HomePlug Green PHY which should be a scaled down version of HomePlug AV.}. Below, we summarize the main advantages:

\begin{itemize}
  \item \textit{Ease of upgrade to future versions}: NB-PLC solutions can be easily implemented as ``soft'' modems using a DSP whereas this is not possible with BB-PLC devices or their scaled down versions.

  \item \textit{Worldwide harmonization}: the \textit{only} available band for PLC in the whole world is the CENELEC band as in some countries the use of other frequencies is prohibited.


  \item \textit{Coexistence}: NB-PLC networks would naturally coexist via FDM with BB-PLC networks thus segregating to two different bands the technologies supporting the very different applications of Smart Grid and home-networking.

  \item \textit{Optimized design}: BB-PLC solutions like IEEE 1901 or ITU-T G.hn were not designed for Smart Grid applications but for home networking or Internet access applications only, whereas HDR NB-PLC design targets explicitly Smart Grid applications and requirements.
\end{itemize}

The above advantages are seen with great interest by the utility, automotive and appliance industries whose choices are greatly influenced by the above criteria. Among the above advantages, the ease of upgrade is of paramount importance for utilities as equipment deployed in the field needs to have long obsolescence horizons and the capability of soft upgrades without the necessity of hardware redeployment is of great economic value (smart meters are considered a very long term investment). Certainly, also cost is a fundamental aspect of technology selection, but it is difficult to compare the true cost of a DSP-based solution versus a silicon-based one as many factors contribute to the ultimate cost of a solution, e.g. design cost, man-hours, field testing, manufacturing process, respinning, etc. Even if a DSP-based solutions may sometimes entail a higher cost versus spun silicon, one may also contend that this may be outweighed by other important factors as explained below.

In the utility-to-meter link, available communications technologies are not mature yet from a standardization point of view, since only LDR NB-PLC standards can be considered today as ``frozen.'' On the other hand, HDR NB-PLC standards have not been ratified yet, it may take a few years to standardize and test in the field, and they may also turn out to be different from currently deployed non-standard solutions like PRIME and G3-PLC. Furthermore, BB-PLC solutions are either over-designed for many Smart Grid applications (IEEE 1901, ITU-T G.hn full profile), or not yet proven in the field (ITU-T G.hn low complexity profile), or still unproven in the field and also non standard-based (HomePlug Green PHY). Thus, availability of a DSP based solution may have several advantages when compared to the ``locked-in'' solution provided by a silicon-based implementation. For the HAN environment one would also probably favor DSP-based solutions since HAN technologies shift and change at a faster rate than what is typically under direct control of the utility. Interestingly, this connectivity uncertainty in the HAN environment may also cause a loss of interest in DSM/DR architectures involving direct load control via the Smart Meter from the utility side and thus giving a growing role to third party energy service providers (cloud-hosted energy management services).

For the above reasons, NB-PLC exhibits very interesting advantages for appliances, meters, and PHEVs - a set of Smart Grid actors that would greatly benefit by direct connectivity with each other. If the industry converges on NB-PLC technologies for these Smart Grid applications, there would be the added advantage of being able to rely on a class of technologies that is decoupled from those BB-PLC technologies that take care of the traditional home networking and Internet access applications. Furthermore, added value services can be easily provisioned by bridging these two networks in a gateway or HEMS.

NB-PLC solutions also have disadvantages with respect to BB-PLC ones when the current rush to deploy equipment in the field is taken into consideration. HDR NB-PLC solutions such as PRIME and G3-PLC have just come out and further validation in the field of these technologies and their effective range and throughput is certainly needed. Similarly, standardization efforts in ITU (G.hnem) and IEEE (1901.2) are not complete yet and further field validation would be needed before proceeding with massive deployments. Also, NB-PLC technologies offer data rates of several kbps (LDR) or at most up to 500 kbps (HDR), and there is a concern that in the long term higher throughput would be required to fulfill the evolution of Smart Grid applications. These concerns seem today to perpetuate the costly paradigm of over-provisioning and have not yet been supported by any quantitative analysis as an accurate estimation of what is really needed for applications close to the load is still an open problem. Thus, a clear justification on why much higher data rates may be needed is still missing - especially when considering that the largest AMI/DR system with direct load control in the world has been operating for the last twenty years using a low data rate UNB-PLC solution (TWACS). Finally, any realistic estimate would also have to take into account the high correlation of the data being generated which calls for smarter sensor aggregation techniques \cite{scaglione_servetto_2003, cristescu} (see also Sect. \ref{sec.ControlSensorNetworking}).

\section{Deployment Aspects: Channel Modeling and Network Topology} \label{sec.DeploymentAspects}
The PLC channel is a very harsh and noisy transmission medium that is difficult to model \cite{Big2003}: it is frequency-selective, time-varying, and is impaired by colored background noise and impulsive noise. Additionally, the structure of the grid differs from country to country and also within a country and the same applies for indoor wiring practices.

Every section of the grid has its own channel characteristics from a communications point of view. On the transmission side, attenuation and dispersion are very small and can be well coped with. However, as we move towards the distribution side and towards the home, attenuation and dispersion grow considerably especially at higher frequencies (see Table \ref{tab.PathLoss}) \cite{Big2003, GotRapDos2004}. On the other hand, the PLC channel is characterized by a high noise level at all voltage levels.

There are various kinds of noises, which are often time, frequency and weather dependent. Furthermore, the PL itself is a noise source. In HV/MV networks, background noise is mainly caused by leakage or discharge events, power converters, transformer, etc. There are also impulsive events due to switching transients, lightening and other discharging events. In the LV and HAN environment, appliances become the cause of a Linear and Periodically Time-Varying behavior of the channel impulse response as well as sources of cyclostationary noise. The Linear and Periodically Time-Varying behavior is due to the fact that the electrical devices plugged in outlets (loads) contain non-linear elements such as diodes and transistors that, relative to the small and rapidly changing communication signals, appear as a resistance biased by the AC mains voltage. The periodically changing AC signal swings the devices over different regions of their non-linear I/V curve and this induces a periodically time-varying change of their resistance. The overall impedance appears as a shunt impedance across the hot and return wires and, since its time variability is due to the periodic AC mains waveform, it is naturally periodic \cite{GallScagXXlptv}. Furthermore, electrical devices are noise generators and, in view of Nyquist theorem, noise also appears to be cyclostationary. Finally, PLs are also both source and victims of electro-magnetic interference so that narrowband noise is also often present. Modem design is thus challenging, especially in dealing with the various sources of noise - which probably represents the most challenging problem in PLC.

In the deployment of Smart Grid devices, and PLC sensors in particular, it is important to devise network planning tools to establish coverage. A first key ingredient is to have accurate and flexible channel modeling tools, especially statical ones. A second element is a network model based on topological properties of the PL network that serves the dual purpose of clarifying the structure of the Smart Grid data source as well as the physical data delivery infrastructure - which in both cases is the grid itself.

\subsection{Recent Advances in Channel Modeling} \label{subsec.ChannelModels}
The issue of channel modeling is of paramount importance as any sensible communications system design must be matched to the particular characteristic of the channel. In particular, the lack of a commonly agreed upon model for the PLC channel has probably slowed down transceiver optimization and the pursuit of general results \cite{Big2003}.

Many authors have been on a quest for a better understanding of the general properties of the PL point-to-point link. Among the advances reported in the last decade, we point the most prominent ones:

\begin{itemize}
\item The multipath law \cite{ZimDos2002mp}.
\item The classification of the several types of noise and their modeling \cite{ZimDos2002emc, katayama:2006}.
\item The isotropy of the PLC channel \cite{GalBanPII2005}.
\item The Linear and Periodically Time-Varying nature of the PLC channel \cite{CanCorDie2006}.
\item The relationship between grounded and ungrounded links, which now can be analyzed under the same formalism  \cite{GalBan2006}.
\item The log-normal distribution of channel attenuation and RMS delay spread of the channel \cite{Galli2009lognormal}.
\item The recent proof that block models similar to those used in wireless and wireline DSL channels can be used in the PLC context as well - an important result since key advances in BB wireless and DSL technologies were fostered by utilizing block transmission models and precoding strategies \cite{GallScagXXlptv}.
\end{itemize}

Most of recent results are related to the BB case and were motivated by the IEEE 1901 and ITU-T G.hn projects. Now that ITU-T G.hnem and IEEE 1901.2 are targeting HDR NB-PLC technologies in the CENELEC/FCC/ARIB bands, more attention will be given to a statistical characterization of these bands and of the through-transformer characteristics. The availability of statistical channel models will aid in gaining a better understanding of the range and coverage that PLC solutions can achieve, a necessary prerequisite when deploying Smart Grid equipment in the field.

We also remark that a network scientific approach, similar to that outlined in Section \ref{sec.Zhifang}, would be needed to provide a truly meaningful statistical model that can guide a large scale deployment. In the next subsections, we will review the latest results in PLC channel modeling.

\subsubsection{Deterministic Models}
At first, PLC channel modeling attempts were mostly empirical and not necessarily tied to PLs per se. The first popular model that attempted to give a phenomenological description of the physics behind signal propagation over PLs is the multipath-model introduced in \cite{Barnes1998, philipps:99, ZimDos1999isplc, ZimDos2002mp}. According to this model, signal propagation along PL cables is predominantly affected by multipath effects arising from the presence of several branches and impedance mismatches that cause multiple reflections. In this approach, the model parameters (delay, attenuation, number of paths, etc.) are fitted via measurements. The disadvantage of this approach is that it is not tied to the physical parameters of the channel. Furthermore, this approach is not even tied to the PLC channel per se as it describes generic signal propagation along any TL-based channel, e.g. see \cite{GalWar02} for twisted pairs.

To overcome this drawback, classical two-conductor TL-theory can be used to derive analytically the multipath model parameters under the assumption that the link topology is known a priori \cite{MengChenGuan2002}. Unfortunately, the computational complexity of this method grows with the number of discontinuities and may become very high for the in-home case (see, for example, Sect. III.A in \cite{GalBan2006}). For this reason, contributions have recently been focusing on frequency domain deterministic models based on TL-theory \cite{BanGal2001, SartDelo2001, EsmaKsch2003, MengChenGuan2004, BanGalPI2005, GalBanPII2005, GalBan2006, AmirKave2006}.

TL-based channel models have today reached a good degree of sophistication as they have been extended to include the multi-conductor TL (MTL) case. Pioneering work on the application of MTL theory to power distribution networks was made by Wedepohl in 1963 \cite{Wede1963mtl}, and tools on mode decoupling were successively introduced by Paul \cite{Paul1996decoupling}. Building on these results, a model for including grounding in LV indoor models was proposed \cite{BanGalPI2005, GalBanPII2005, GalBan2006}. The MTL approach is a natural extension of the two-conductor modeling to include the presence of additional wires, such as the ground wire and allows to compute the transfer function of both grounded and ungrounded PL links by using transmission matrices only. These results allow us to treat with the same formalism both grounded and ungrounded indoor PLC channels.
As an example, let us consider a generic topology of a PL link between two devices located at nodes X and Y as shown in Figure \ref{fig.PLC_Link_Equivalent}. If the PL link is not grounded, then the corresponding topology is amenable of simple two-conductor TL theory description via two-port networks. If grounding is present at the main panel, a mirror topology representing what is referred to as the ``companion model'' must be added as a bridged tap located at the main panel as shown in Figure \ref{fig.PLC_Link_Equivalent}.

\insertonefig{8.8cm}{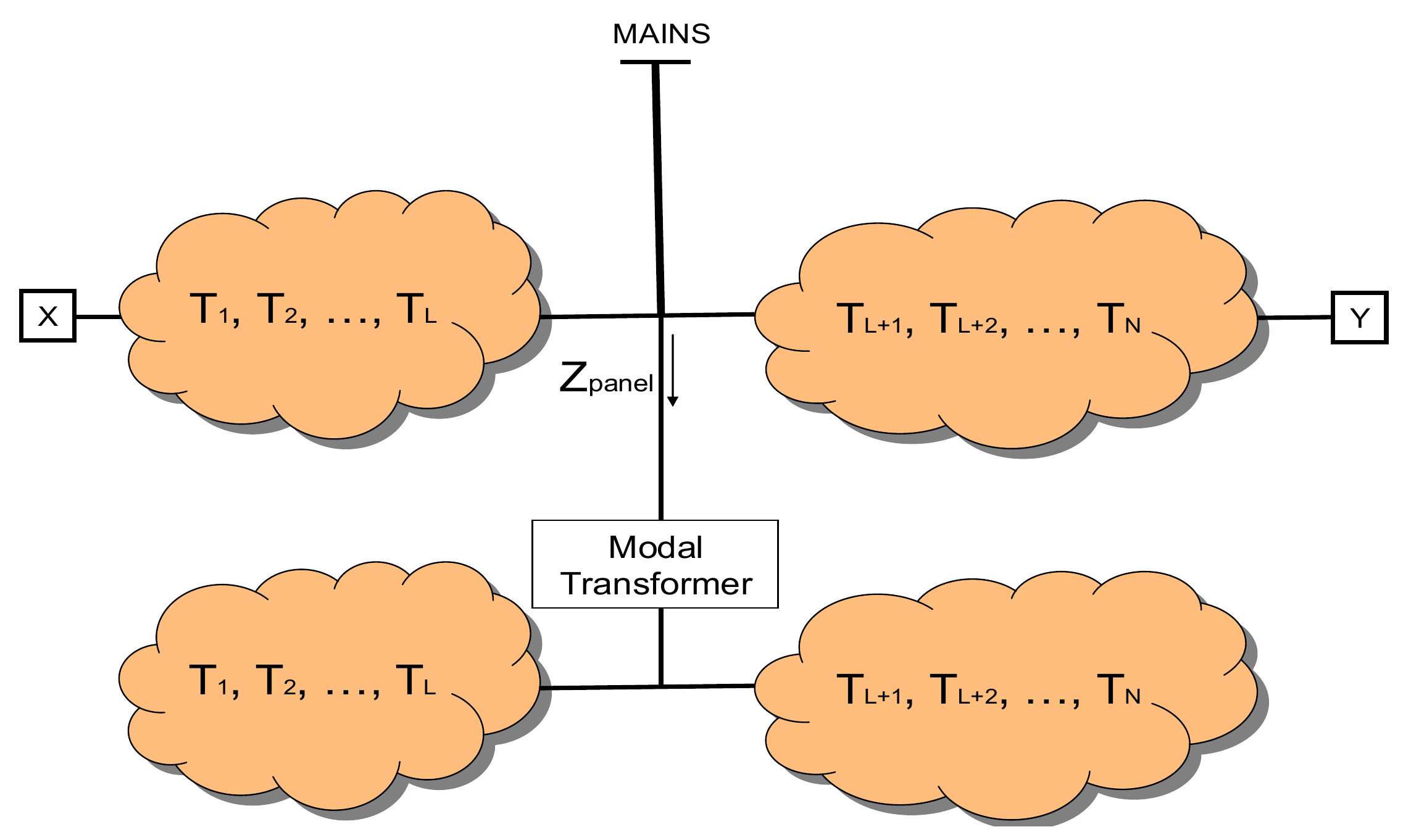}{\label{fig.PLC_Link_Equivalent}}
{The equivalent power line link in terms of cascaded two-port networks when grounding is present.}

\subsubsection{Statistical Models}
The transfer function of a TL-based channels can be deterministically calculated once the link topology is known. However, the variability of link topologies and wiring practices give rise to a stochastic aspect of TL-based channels that has been only recently addressed in the literature. To encompass several potential scenarios and study the coverage and expected transmission rates of PLC networks, one needs to combine these MTL-based deterministic models with a set of topologies that are representative of the majority of cases found in the field. This approach is reminiscent of what has been done in xDSL context with the definition of the ANSI and CSA loops. Although this approach may be suitable for the outdoor MV/LV cases, its applicability to the in-home case may be questionable due to the wide variability of wiring and grounding practices.

An excellent approach to the generation of random in-home topologies was made by Esmailian et al. \cite{EsmaKsch2003}, where the US National Electric Code was used to set constraints on the topologies in terms of number of outlets per branch, wire gauges, inter-outlet spacing etc. This is probably the most realistic and accurate way of generating randomly channel realizations, although a generalization of this approach requires the knowledge of the electric codes of every country. Only a few other attempts have been made to develop a statistical model for the PLC channel, e.g. \cite{Tonello2007}, \cite{TlicZeddMoulTPDI2008}.

A useful result for the modeling of the PLC channel and the calculation of its achievable throughput was the discovery that attenuation in LV/MV PLC channels is log-normally distributed \cite{Galli2009lognormal, Galli2010TwoTaps, Link:Galli2011wirelinemodel}. Considering signal propagation along TLs as multipath-based, channel distortion is present at the receiver due not only to the low pass behavior of the cable but also to the arrival of multiple echoes caused by successive reflections of the propagating signal generated by mismatched terminations and impedance discontinuities along the line. This is a general behavior and is independent of the link topology or, in the case of PLs, of the presence of grounding \cite{GalBan2006}. According to this model, the transfer function is \cite{ZimDos2002mp}:
\begin{equation}\label{eq.Hfmp}
    H(f) = \sum_{i=0}^{N_{paths}-1} g_i(f) e^{-\alpha (f) v_p \theta_i} e^{-j2 \pi f \theta_i}
\end{equation}
where $g_i(f)$ is a complex number generally frequency dependent that depends on the topology of the link, $\alpha (f)$ is the attenuation coefficient which takes into account both skin effect and dielectric loss, $\theta_i$  is the delay associated with the $i$-th path, $v_p$ is velocity of propagation along the PL cable, and $N_{paths}$ is the number of non-negligible paths. Similarly, we can write in the time domain:
\begin{equation}\label{eq.htmp}
    h(t) = \sum_{i=0}^{N_{paths}-1} e_{ep}^{(i)}(t-\theta_i)
\end{equation}
where $e_{ep}^{(i)}(t)=FT^{-1} \left[ g_i(f) e^{-\alpha (f) v_p \theta_i} \right]$ is the signal propagating along the $i$-th path and its amplitude and shape are a function of the reflection coefficients $\rho^{(i)}$ and the transmission coefficients $\xi^{(i)}=(1+\rho^{(i)})$ associated to all the impedance discontinuities encountered along the $i$-th path, and of the low-pass behavior of the channel in the absence of multipath (for analytical expressions of $\rho^{(i)}$ and $\xi^{(i)}$, see \cite{GalBan2006} for the case of forward traveling signal paths and \cite{GalWar02} for the case of backward traveling echo paths). Thus, the path amplitudes are a function of a cascade (product) of several random propagation effects and this is a condition that leads to log-normality in the central limit since the logarithm of a product of random terms becomes the summation of many random terms. Since log-normality is preserved under power, path gains are log-normally distributed as well. Finally, since the sum of independent or correlated log-normal random variables is well approximated by another log-normal distribution \cite{MehMolWu06}, we can finally state that also the PLC channel average gain (or attenuation) is log-normally distributed.

Empirical confirmation of this property of the PLC channel has been reported for indoor US sub-urban homes \cite{Galli2009lognormal}, indoor US urban multiple dwelling units \cite{Galli2010TwoTaps}, and for US outdoor MV underground PLs \cite{Link:Galli2011wirelinemodel}. Furthermore, these PLC channel characteristics have also been observed in other wireline channels such as coax and phone lines, so that a new generalized wireline statistical channel model has been recently formulated in \cite{Link:Galli2011wirelinemodel}. The availability of these results greatly facilitates the study of coverage which is necessary for proper planning and deployment.

\subsection{Topological Analysis of Power Grids} \label{sec.Zhifang}

The study of the topology and electrical characteristics of the power grid provides a two major benefit: (1) it provides a deep understanding of the network dynamics, hence the information traffic in the PLC-based network; (2) it complements fading channel models by including topological aspects that affect a PLC-based network.

Recent results on the topological characteristics of the transmission side of the power grid were reported in \cite{WangScaglioneThomas_SGpowergridtopo:2010}. These results show that the transmission power grid topology has sparse connectivity, well emulated by a collection of subgraphs connected in a ring, each of which closely matches
the characteristics of a small-world network \cite{watts98}. A key observation that follows is that the path distances separating nodes are relatively small when compared to the size of the network, which clearly has beneficial implications on the communication delay if the topology of the power delivery network matches the communication one - the case when PLC is used. Peculiarities in this section of the grid include also the exponential tail of the nodal degree distribution and the heavy tail distribution of the line impedances.

In this paper we extend the analysis in \cite{WangScaglioneThomas_SGpowergridtopo:2010} to the distribution side of the power grid and we report new results based on a sample 396-node MV distribution network which comes from a real-world US distribution utility mainly located in a rural area. This is a first step in achieving a better understanding of the topological characteristics of the distribution network and its implications on the use of PLC.

\insertonefig{8.8cm}{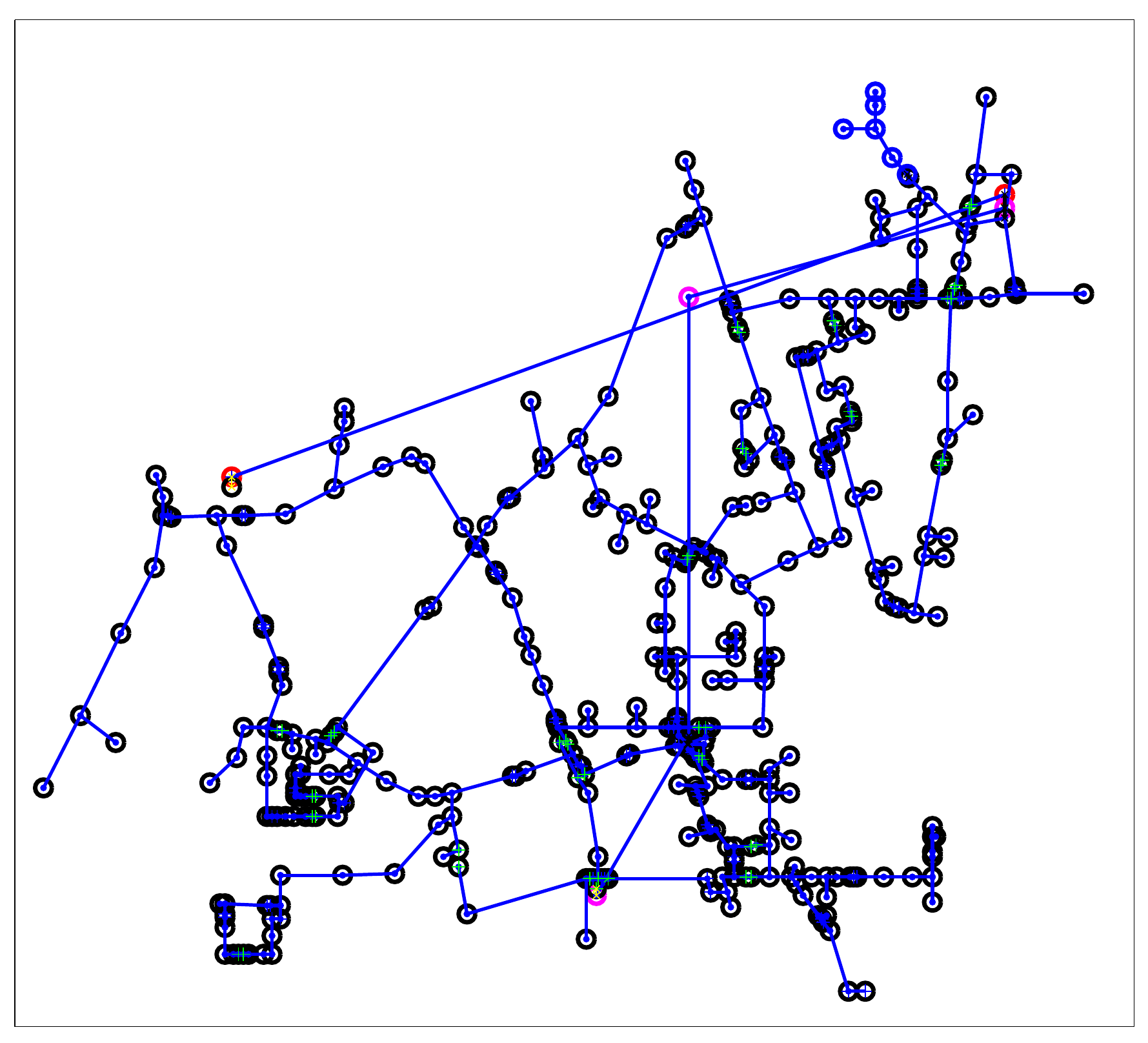}{\label{fig.NYSEG_topo}}
{A 396-node MV distribution network in a rural area of the US. Components: bus (circle), line branches (line ending with dots), switches (line ending with `+'s), transformers (lines ending with `x's), open or out of service component (green dotted line); the node color representing its voltage levels: 115 kV (red), 34.5 kV(magenta), 12.47 kV(black), 4.80 kV(blue).}

\subsubsection{Structure of Distribution Network}

A distribution network carries electricity from the transmission system and delivers it to end users. Typically, the network would include MV (less than 50 kV) lines, electrical substations and pole-mounted transformers, LV (less than 1 kV) distribution wiring and sometimes electricity meters.

In the low and medium voltage sections of the grid the physical layout is often restricted by what land is available and its geology. The logical topology can vary depending on the constraints of budget, requirements for system reliability, and the load and generation characteristics. Generally speaking, there are a few typical kinds of topology in the distribution network: ring, radial or interconnected.

A radial network is the cheapest and simplest topology for a distribution grid, and the one more often encountered.  This network has a tree shape where power from a large supply radiates out into progressively lower voltage lines until the destination homes and businesses are reached. It is typical of long rural lines with isolated load areas. Today's grid is radially operated with respect to the current transmission system, but this topology will not hold anymore when DER will be integrated into the grid. Unfortunately this topology is the worst in terms of maximum communication delay because the number of hops between its nodes tend to grow in the order of the size of the network.

An interconnected or ring network is generally found in more urban areas and will have multiple connections to other points of supply. These points of connection are normally open but allow various configurations by the operating utility by closing and opening switches. Operation of these switches may be by remote control from a control center or by a lineman. The benefit of the interconnected model is that, in the event of a fault or a required maintenance, a small area of the network can be isolated and the remainder kept on supply.

Most areas provide three phase industrial service. A ground is normally provided, connected to conductive cases and other safety equipment, to keep current away from equipment and people. Distribution voltages vary depending on customer need, equipment and availability. Within these networks there may be a mix of overhead line construction utilizing traditional utility poles and wires and, increasingly, underground construction with cables and indoor or cabinet substations. However, underground distribution is significantly more expensive than overhead construction. Distribution feeders emanating from a substation are generally controlled by a circuit breaker which will open when a fault is detected. Automatic circuit reclosers may be installed to further segregate the feeder thus minimizing the impact of faults. Long feeders experience voltage drop requiring capacitors or voltage regulators to be installed. However, if DSM is successful and peak demand per customer is reduced, then longer feeders can be tolerated and included in the design phase provided that demand peaks can still be deterministically bounded when DSM/DR applications are running - note that this may entail regulatory intervention to mandate some form of predictability in customer behavior.

\subsubsection{Graph Theoretic Analysis of a Sample MV Distribution Network}
The logical topology of the sample 396-node MV distribution network analyzed here is shown in Figure \ref{fig.NYSEG_topo}. The power supply comes from the 115 kV-34.5 kV step-down substation. Most nodes or buses in the network are 12.47 kV, and only a small number of them are 34.5 kV or 4.8 kV.

As shown in Figure \ref{fig.NYSEG_topo}, an MV network usually comprises different voltage levels, separated by transformers. As mentioned in Sect. \ref{sec.AMR-AMI}, there is  not enough evidence to characterize statistically the through-transformer behavior of NB-PLC signals. Thus, in the analysis of the graph properties of the distribution network, one would have to consider two extreme cases: 1) all transformers block PLC signals; 2) all transformers allow PLC signals through. The two cases become the same if appropriate couplers are installed in order to bypass transformers and obtain system-wide connectivity.

In the following topology analysis of the sample MV network, it is assumed that wireless or wired couplers have been implemented at the locations of transformers and switches, so that the network connectivity will not be affected by transformer types or switch status. On the other hand, if couplers are missing, the network will be segmented into several sections either by the transformers or by the open switches. For the sample MV network analyzed here, most buses ($>95\%$) in the network are at the same voltage level of 12.47 kV. Therefore the topology analysis result of the separated 12.47 kV subnetwork is in fact very close to that of the whole connected graph.

The topology metrics we evaluated include the following:
\begin{itemize}
  \item $(N,m)$: the total number of nodes and branches, which well represents the network size.
  \item $\langle k \rangle$: the average node degree, which represents the average number of branches a node connects to.
  \item $\langle l \rangle$: the average shortest path length in hops between any pair of nodes.
  \item $\rho$: the Pearson correlation coefficient, which evaluates the correlation of node degrees in the network. This measure reflects if a node adjacent to a highly connected node has also a large node degree.
  \item $\lambda_2(L)$: the algebraic connectivity, which is the second smallest eigenvalue of the Laplacian matrix and is an index of how well a network is connected and how fast information data can be shared across the network.
  \item $C(G)$: the clustering coefficient, which assesses the ratio of nodes tending to cluster together.
\end{itemize}

The Laplacian matrix is a matrix representation of a graph \cite{Book:Chung1997GraphTheory}. For an $n$-node simple network without self-loops and duplicate links, the Laplacian $L:=(l_{i,j})_{n\times n}$ is defined as: $l_{i,i} = deg(\textrm{node}_i)$; for $i\ne j$, $l_{i,j}=-1$, if $\textrm{node}_i$ is adjacent to $\textrm{node}_j$, otherwise $l_{i,j}=0$.

The result of the analysis is listed in Table \ref{tab:topometrics} with comparison to other two transmission networks: the IEEE-300 system represents a synthesized network from the New England power system and has a comparable network size as the 396-node MV distribution network we analyzed; the Western System Coordinating Council (WSCC) grid is the interconnected system of transmission lines spanning the Western United States plus parts of Canada and Mexico and contains 4941 nodes and 6594 transmission lines. It is well known that transmission and distribution topologies differ, nevertheless we decided to comment on these differences in a quantitative manner as this exercise is useful for several reasons. For example, it allows us to better understand the characteristics of the transmission and distribution networks as information sources; it allows us to optimize the design of the distribution PMU based WAMS rather than attempting to duplicate the existing transmission one which is tailored to a network with very different topological characteristics; it can tells us how the distribution topology can be ``modified'' to achieve some advantageous characteristics of the transmission network, i.e. shorter path lengths between nodes, better algebraic connectivity, etc.

\begin{table*}[!t]
\renewcommand{\arraystretch}{1.3}
\caption{Topological Characteristics of typical Transmission Networks and the sample MV Distribution Network analyzed here.}
\label{tab:topometrics}
\centering
\begin{tabular}{l|c|c|c|c|c|c}
\hline \hline
{} & {$(N,m)$} & {$\langle k \rangle$} & {$\langle l \rangle$} & {$\rho$} & {$\lambda_2(L)$} & {$C(G)$}\\
\hline
{IEEE-300} & (300, 409) & 2.73 &9.94 & 	-0.2206	& 0.0094 & 0.0856\\
\hline
{WSCC} &  (4941, 6594) & 2.67 &18.70 &  0.0035 & 0.00076 & 0.0801  \\
\hline
{396-node MV-Distr} &  (396, 420) & 2.12 & 21.10 & -0.2257 & 0.00030 & 0              \\
\hline \hline
\end{tabular}
\end{table*}

From Table \ref{tab:topometrics} we can see that the 396-node MV distribution network has an average node degree of $\langle k \rangle = 2.12$, which is comparable to, although a little bit lower than, that of  the other two transmission networks, the IEEE-300 system and the WSCC system. That means its average connecting sparsity is about at the same level as the compared transmission networks. However, the sample MV distribution network has a much longer average path length of $\langle l \rangle = 21.10$ in hops than the IEEE-300 system and, interestingly, it is even longer than that of the much larger 4941-node WSCC system. More specifically, any node in this MV distribution network is about 16.50 hops away from node-$1$ or node-$2$ which are $115$-KV buses at the HV side of the two step-down supply transformers and may likely serve as the traffic sinks.

Looking at the algebraic connectivity $\lambda_2(L)$, the 396-node MV distribution network has a much weaker overall connectivity compared to the transmission networks, i.e. $\lambda_2(L) = 0.00030$ versus 0.0094 (IEEE-300) and 0.00076 (WSCC). This result shows that this topology is highly prone to become a disconnected graph under node failure (islanding). Finally, the most distinctive difference we found lies in the fact that the 396-node MV distribution network has a clustering coefficient equal to zero, compared to the clustering coefficient of 0.0856 for the IEEE-300 system and 0.0801 for the WSCC system. This means that no node in the sample MV distribution network is the vertex of a complete subgraph (triangle). MV distribution networks not located in rural areas are generally less prone to becoming a disconnected graph as in urban areas it is not unusual that utilities provide link redundancy, e.g. adding rings. If the distribution network becomes a disconnected graph, data connectivity obviously suffers if PLC is used. This vulnerability of the distribution network can be alleviated by adding judiciously wireless links to complement the PLC based network with the goal of improving network connectivity as well as shortest path lengths characteristics. Thus, the realization of a hybrid PLC/wireless infrastructure that exploits synergistically the strengths of PLC and wireless could drastically improve the robustness and reliability of the data network in the distribution grid and also add self-healing capabilities. It is then convenient to split the hybrid network so obtained into relatively independent and smaller layer 3 clusters. As suggested in \cite{GalLusSuc05}, this can be accomplished using a two-step approach based on Graph Partitioning that yields to a robust network design characterized by balanced domains with minimal inter-domain traffic.

\insertonefig{8.8cm}{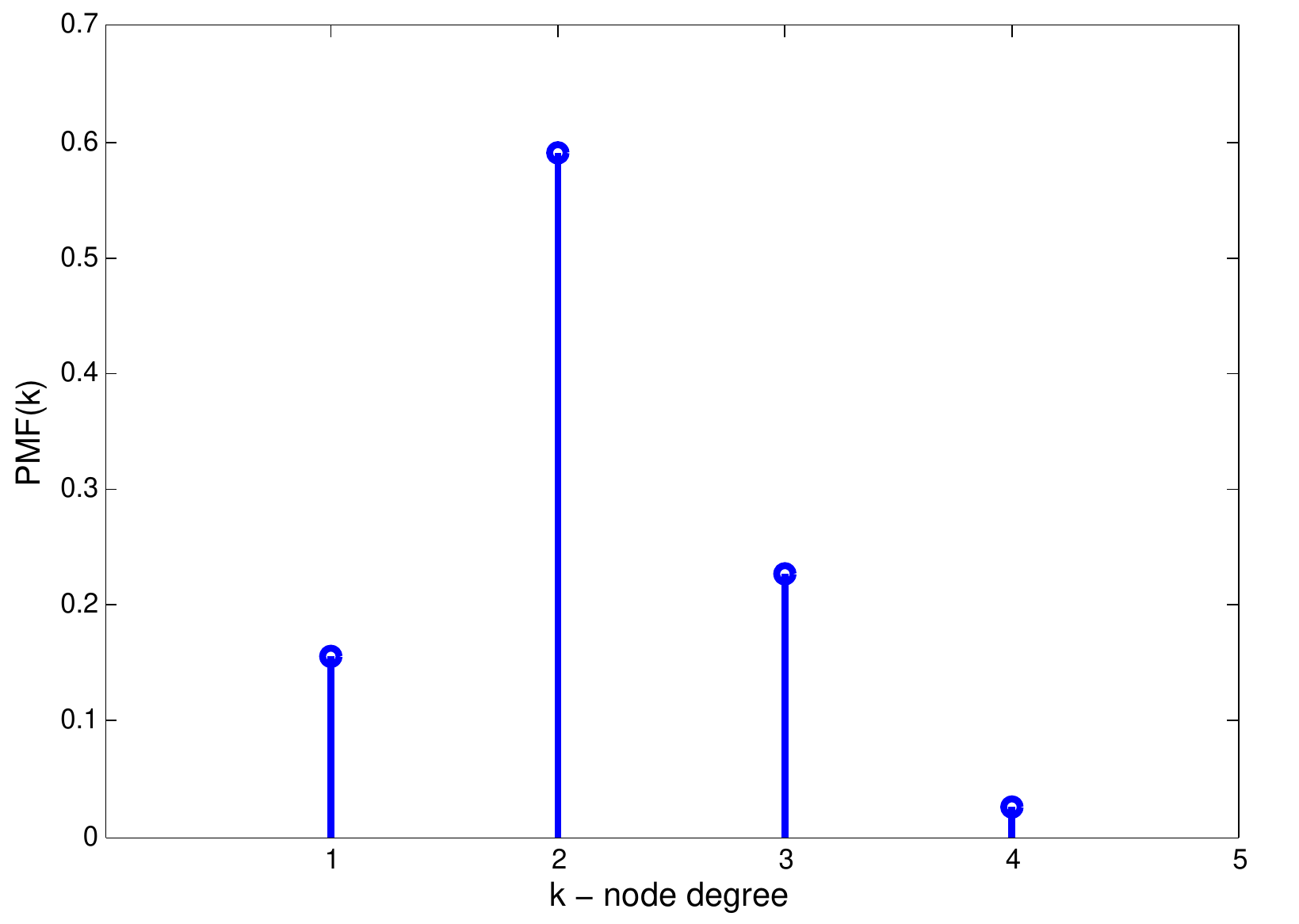}{\label{fig.NYSEG_nodedeg}}
{The probability mass function of the node degrees in the sample 396-node MV distribution network.}

As we have learned from \cite{WangScaglioneThomas_SGpowergridtopo:2010}, the average node degree of a power grid transmission network tends to be quite low and does not scale as the network size increases. The topology of a transmission network has salient \emph{small-world} properties \cite{watts98}, since it features a much shorter average path length (in hops) and a much higher clustering coefficient than that of Erd$\ddot{o}$s-R$\acute{e}$nyi random graphs with the same network size and sparsity. While small-world features have been recently confirmed for the HV transmission network \cite{WangScaglioneThomas_SGpowergridtopo:2010}, the sample MV network used here implies that a power grid distribution network has a very different kind of topology than that of a HV network and obviously it is not a \emph{small-world} topology.

\insertonefig{8.8cm}{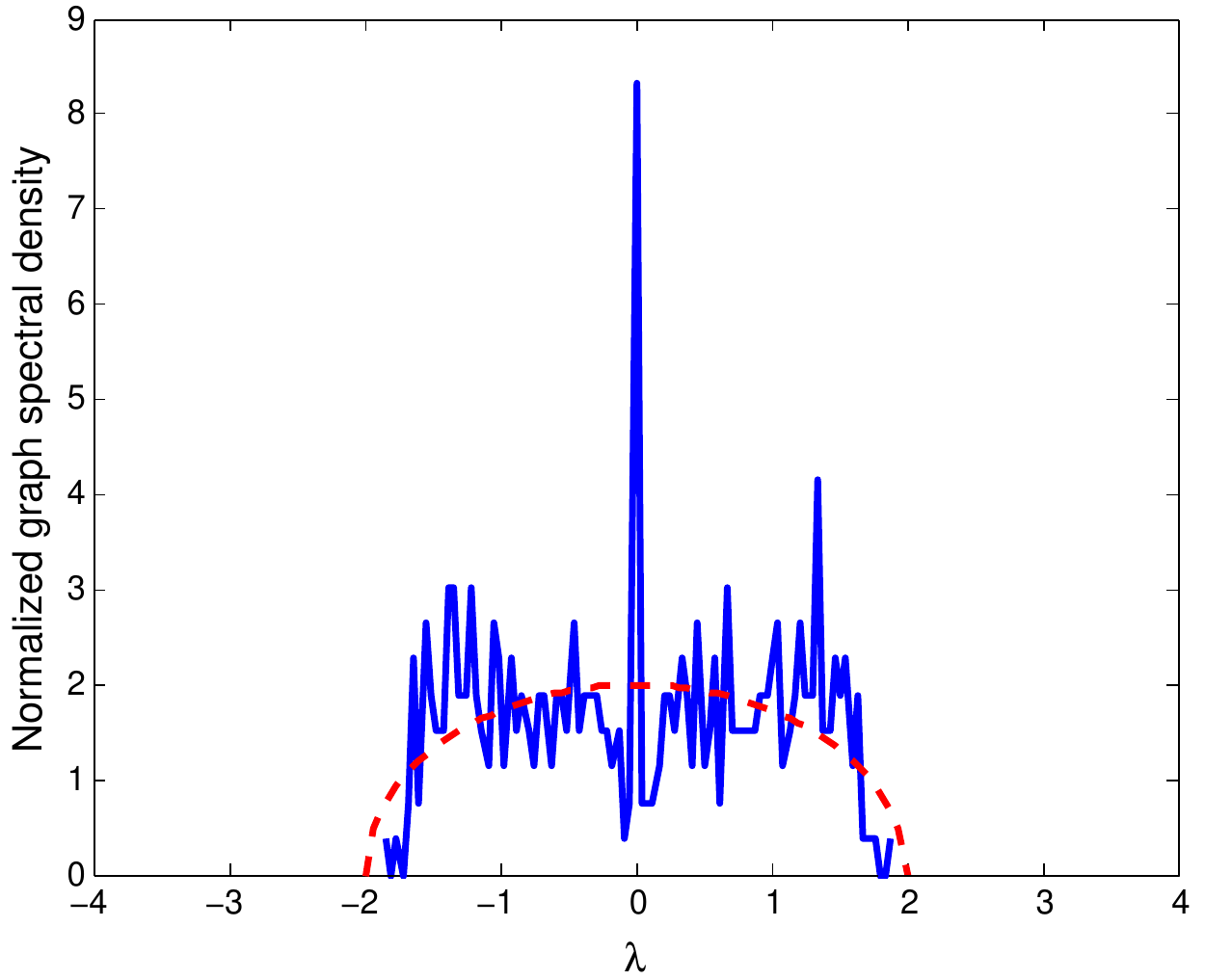}{\label{fig.NYSEG_graphD}}
{The normalized graph spectral density of the sample 396-node MV distribution network, $\widetilde{\rho}(\lambda)$ vs. $\widetilde{\lambda}$: the dotted line of semi-circle represents the graph spectral density of random graph networks.}

The node degree distribution of the 396-node MV distribution network is shown in Figure \ref{fig.NYSEG_nodedeg}. The maximum node degree in the network equals to 4 - which is much smaller than what is found in the transmission side of the grid where maximum nodal degrees of 20 or 30 can be found.  The Figure shows that about 16\% of the nodes connect to only one branch,  60\% connect with 2 branches, 22\% with 3 branches, and only 2\% with 4 branches.

Figure \ref{fig.NYSEG_graphD} depicts the network's spectral density, which is a normalized spectral distribution of the eigenvalues of its adjacency matrix. The spectrum of an Erd$\ddot{o}$s-R$\acute{e}$nyi random graph network, which has uncorrelated node degrees, converges to a semicircular distribution (see the semi-circle dotted line on the background in Figure \ref{fig.NYSEG_graphD}). According to \cite{FarkasDerenyi_GraphSpectraRealWorldNetwork:2001}, the spectra of real-world networks have specific features depending on the details of the corresponding models. In particular, scale-free graphs develop a triangle-like spectral density with a power-law tail; whereas a small-world network has a complex spectral density consisting of several sharp peaks. The plot in Figure \ref{fig.NYSEG_graphD} indicates that the sample MV distribution network is neither a scale-free network nor a small-world network.

We also analyze the branch lengths in the MV distribution network. The corresponding probability mass function is shown in Figure \ref{fig.NYSEG_linelength}. It indicates that most of the branches are shorter than 1,067 m (3,500 ft) and the branch length distribution has an exponential tail with only a very small number of branches of extremely long length.

\insertonefig{8.8cm}{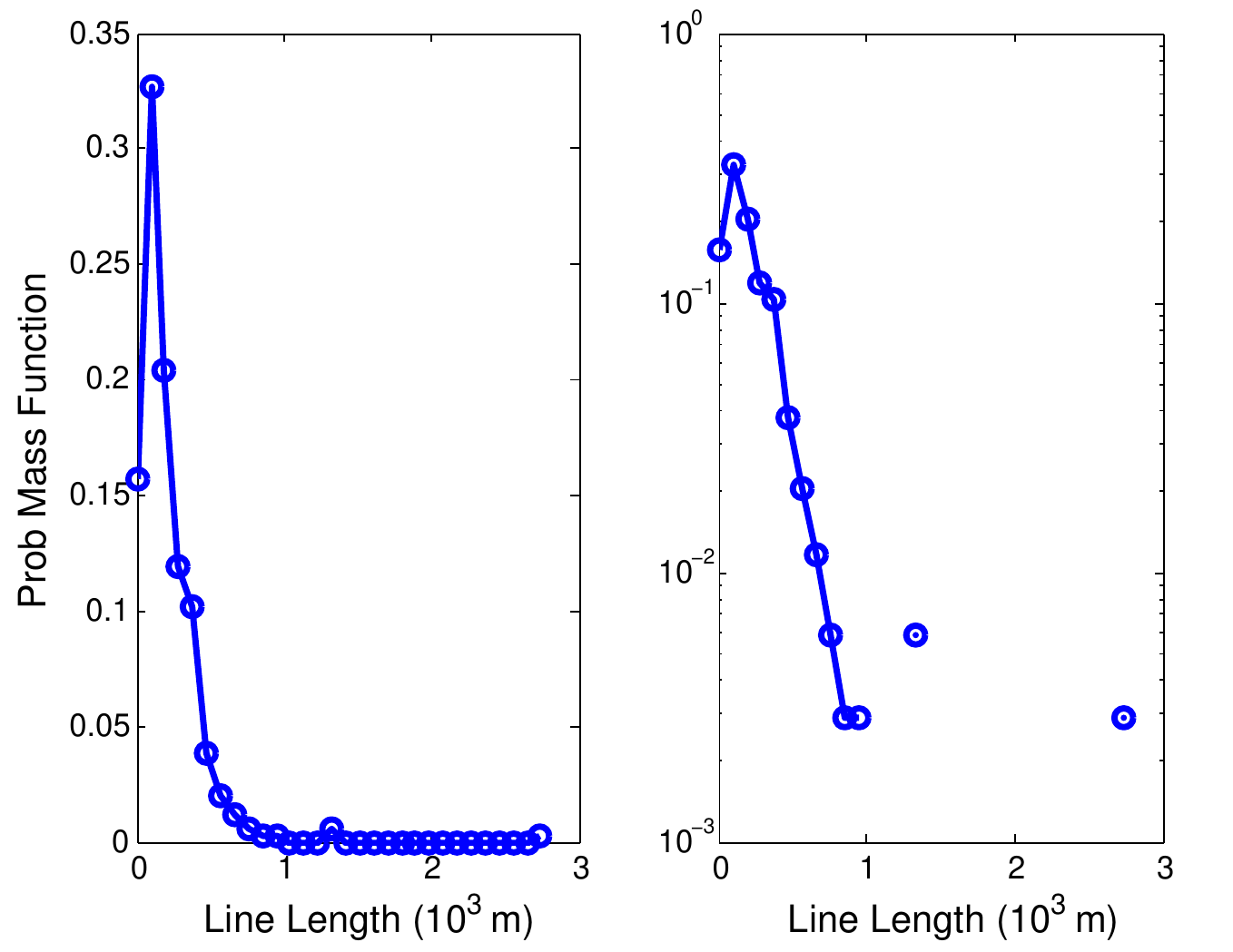}{\label{fig.NYSEG_linelength}}
{The probability mass function of the line length in the sample 396-node MV distribution network: (left) probability versus length; (right) log-probability versus length, where the existence of an exponential trend in the tail is clearly visible.}

\subsection{The LV Distribution Network} \label{subsec.ZhifangLV}

It is difficult to obtain example data about LV distribution network topologies. Generally speaking, an LV distribution network is radial, and has a similar network topology as an MV distribution network except that it may have more nodes with shorter branch length.

\section{Final Considerations and Recommendations} \label{sec.Conclusions}
We conclude this paper by making some final considerations on the advantages of PLC for utility applications and also making some recommendations on the methodological aspects of Smart Grid design. Some of these recommendations transcend PLC or any specific Smart Grid communications technology. The reason for this is that there are still many open problems related to the implementation of the Smart Grid so that the most pressing aspect today is to determine the right methodological approach rather than giving prematurely specific recommendations on networking technologies. At the same time, we hope to leave the reader with a hint of our optimism for the use of PLC in the Smart Grid.

\subsection{Final Considerations on PLC for the Smart Grid}

As this paper has shown, there are many applications scenarios in the Smart Grid that require a diversity of communications technologies. Although it is expected that the Smart Grid will be supported by an heterogeneous set of networking technologies, we hope to have shown that PLC is an excellent and mature technology that can support a wide variety of applications from the transmission side to the distribution side and also to and within the home. As also pointed out in Sect. \ref{subsec.BestPLC}, there are many PLC technologies either already available or currently under standardization and it is very important to refrain from advocating a single PLC technology rather than exercising a judicious choice in the selection of the right PLC technology for the right set of applications. There are many possible choices at the disposal of communications and utility engineers, and the vast majority of these technologies can find a suitable application within the Smart Grid.

Many of the available PLC technologies are well separated in frequency from each other so that a good design of the analog front end would eliminate interference between non-interoperable technologies. However, there are also multiple non-interoperable technologies that operate in the same band so that coexistence mechanisms are required to alleviate the performance degradation due to to mutual interference. Although it may be true as some believe that coexistence stands in the way of interoperability and may delay industry alignment behind a single standard, it is important to understand that the usage of PLC spectrum is not regulated so that any PLC technology can use channel resources without having any legal obligation to protect other PLC technology from interference. Thus, any deployed PLC technology is a source/victim of interference to/from the installed base of PLC devices if a common coexistence mechanism is not supported since there is not enough bandwidth to implement efficiently FDM as in the WiFi and coax cases. Furthermore, the implementation of coexistence in PLC transceivers also allows that diversification of deployment that is today a necessary ingredient for achieving a better understanding of how to build the Smart Grid without having to pay the penalty of interference, performance degradation, and service disruption.

On the basis of the above considerations, coexistence can be seen as a transitory and necessary ``evil'' that will allow the industry to align behind the right PLC technology for the right application on the basis of field deployment data and not on the basis of a pre-selection strategy. Furthermore, coexistence will also ensure that the operation of Smart Grid and home networking devices can be decoupled and allowed to mature at their traditional obsolescence rate - even if operating in the same band. Last but not least, coexistence will also allow utilities and other service providers to avoid having to resolve ``service'' issues caused by the interference between non-interoperable PLC devices supporting different applications.

Finally, we below summarize the fundamental benefits offered by PLC when it is employed for the Smart Grid and, more in general, for utility applications:

\begin{enumerate}

  \item Utility applications almost always require redundancy in protection and control, and the need for redundancy must include the availability of redundant communications channels: PLC allows to exploit the existing wired infrastructure thus greatly reducing the cost of deploying a redundant communications channel.

  \item The use of PLC allows blurring together the traditionally separated functions of \textit{sensing} and \textit{communicating} as a PLC transceiver can be designed to switch between functioning as a ``sensor'' and as a ``modem.''

   \item PLs often represent the most direct route between controllers and IEDs when compared to packet switched public networks, so that PLC offers substantial advantages when dealing with applications such as tele-protection where ensuring a low and bounded latency is crucial.

   \item Power lines provide a communication path that is under the \textit{direct and complete} control of the utility which is a fundamental benefit when operating in countries where telecom markets are de-regulated.

   \item There is a wide variety of PLC technologies that can find a role in most Smart Grid application, so that PLC can indeed provide a wide class of technologies that can be employed as a communications solution from the transmission side of the grid down to the HAN.

\end{enumerate}

\subsection{Architecture Must Come First!}
Utilities, vendors, regulators and other forces are spearheading deployments - especially in AMI. Given that what is put in the field today will be there for some decades, addressing the design aspects well from the beginning is very important. However, getting things right from the onset is complicated because of the current fog surrounding what the Smart Grid architecture should be. A fundamental priority is thus to accelerate the work on the development of an architectural framework that not only maps existing standards to the ultimate vision of what the Smart Grid will be, but also individuates standards gaps that threaten interoperability. In the US, NIST is leading an effort in this direction, trying to lay down a strategy to integrate legacy systems and new Smart Grid technologies with the goal of preserving system interoperability. An international effort aimed at defining a detailed Smart Grid architecture to ensure system interoperability from generation to load is ongoing in the IEEE P2030 standard \cite{Link:P2030}.

While establishing a migration path is a sensible approach, there also has to be some judicious selection of which technologies should be carried to the future, as also John Boot (General Electric) stated in his IEEE ISPLC 2010 keynote \cite{Link:BootISPLC10keynote}: ``\textit{There needs to be an understanding that Smart Grid Standards are forward looking only and that the migration will take perhaps decades until all equipment adheres to new standards. However, we should not try to push old standards into the future or the migration will never take place}.''

\subsection{Avoid the Temptation for a Single Networking Technology}
The pressure of administrations, regulators, and some industry sectors to accelerate the deployment of the Smart Grid has sometimes pushed the collective thinking into making decisions based on two questionable assumptions:

\begin{itemize}
  \item Off the shelf technologies, even if designed and implemented for completely different applications, can be massively and seamlessly utilized in Smart Grid - and this even before fully understanding what the actual requirements for those applications really are.

  \item The choice of a single technology for the implementation of certain Smart Grid applications such as DSM or AMI would accelerate reaping Smart Grid benefits since it would allow the industry to align behind a single common technology - an alignment that has not occurred yet under normal market dynamics.
\end{itemize}

The efforts devoted to the realization of the Smart Grid must take into account that the Smart Grid is, from every point of view, an \textit{on-going experiment} - an experiment that will continue for decades to come. The understanding that the Smart Grid is still an \textit{experiment} should lead us to make choices at this stage that encompass a diversity of solutions and implementations in order to be able to achieve a better understanding of how to cope with the very complex problem of building the Smart Grid.

\subsection{Stability and Blackout Prevention: A Sisyphean Quest?} \label{sec.ConclusionsSisyphus}
There is a very interesting body of published work that uses statistical physics tools (e.g., percolation theory in random geometric graphs \cite{CallDuncNew00}) to analyze ``phase transitions'' with application to blackout analysis \cite{DobsCarrLynch07}. The characterization of these phase transitions and their triggering mechanisms are essential to the analysis of the impact of distributed control algorithms on the overall stability of the grid. Recent analysis of US blackouts found supporting evidence of the validity of a complex dynamics behavior of the power grid \cite{CarrNewmDobs04, ChenThorDobs05}. As stated in \cite{CarrLyncNewm03}, ``\textit{The slow evolution of the power system is driven by a steady increase in electric loading, economic pressures to maximize the use of the grid, and the engineering responses to blackouts that upgrade the system. Mitigation of blackout risk should account for dynamical effects in complex self-organized critical systems. For example, some methods of suppressing small blackouts could ultimately increase the risk of large blackouts.}'' Furthermore, Hines et al. also point out \cite{HineAptTalu08}: ``\textit{Despite efforts to mitigate blackout risk, the data available from the North American Electric Reliability Council (NERC) for 1984-2006 indicate that the frequency of large blackouts in the United States is not decreasing}.''

Blackout data from several countries suggests that the frequency of large blackouts is governed by a power-law, which is consistent with the grid being a complex system designed and operated near a critical point \cite{CarrNewmDobs04}.
Although it is possible that changes to the grid near the load (DSM, DR, DER, etc.) could change the power-law distribution of blackout size, not much is actually known about this. As a consequence, it is difficult today to draw general conclusions on the overall effects that ``smartness'' will have on the stability of the power grid \cite{Link:Galli10smartness}.

\section*{Acknowledgment}
The Authors would like to express their gratitude to John Boot (General Electric), Anand Dabak (Texas Instruments), Mischa Schwartz (Columbia University), and Gary Stuebing (Duke Energy) for many stimulating discussions on the subject of this manuscript. The Authors also owe many thanks to the anonymous reviewers for their constructive criticism.

\bibliographystyle{IEEEtran}
\bibliography{IEEEabrv,All_Papers,PLC_Papers,SmartGrid}

\begin{IEEEbiography}[\addphoto{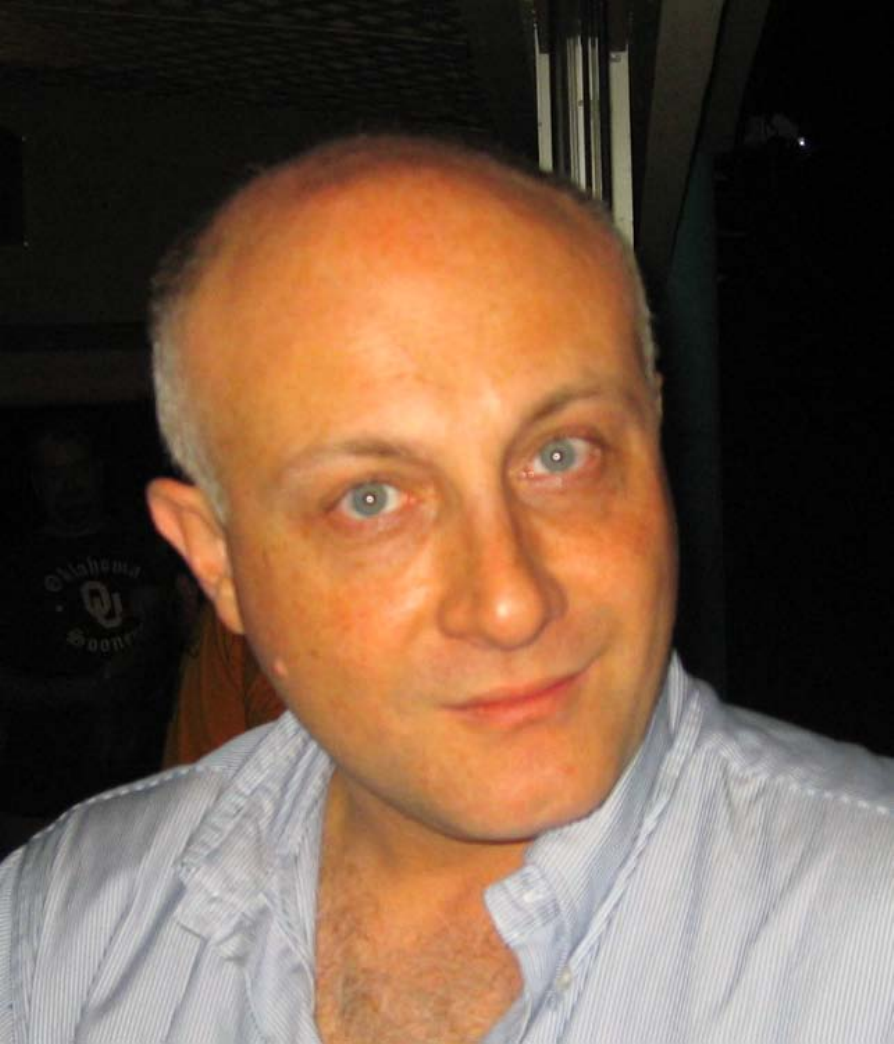}]
{Stefano Galli} (S'95, M'98, SM'05) received his M.Sc. and Ph.D. degrees in Electrical Engineering from the University of Rome ``La Sapienza'' (Italy) in 1994 and 1998, respectively. Currently, he is the Director of Technology Strategy of ASSIA where he leads the overall standardization strategy of the company and also contributes to the company's efforts in the area of wired/wireless access and home area networking. Prior to this position, he was in Panasonic Corporation from 2006 to 2010 as Lead Scientist in the Strategic R\&D Planning Office and then as the Director of Energy Solutions R\&D. From 1998 to 2006, he was a Senior Scientist in Bellcore (now Telcordia Technologies).\\
Dr. Galli is serving as Member-at-Large of the IEEE Communications Society (ComSoc) Board of Governors and is involved in a variety of capacities in Smart Grid activities. He currently serves as: Chair of the PAP 15 Coexistence subgroup instituted by the US National Institute of Standards and Technology (NIST), Chair of the IEEE ComSoc Ad-Hoc Committee on Smart Grid Communications, Member of the Energy and Policy Committee of IEEE-USA, and Editor for the IEEE Transactions on Smart Grid and the IEEE Transactions on Communications (Wireline Systems and Smart Grid Communications). He is also the founder and first Chair of the IEEE ComSoc Technical Committee on Power Line Communications (2004-2010), and the past Co-Chair of the ``Communications Technology'' Task Force of the IEEE P2030 Smart Grid Interoperability Standard (2009-2010).\\
Dr. Galli has worked on a variety of wireless/wired communications technologies, is an IEEE Senior Member, holds fifteen issued and pending patents, has published over 90 peer-reviewed papers, has submitted numerous standards contributions, and has received the 2010 IEEE ISPLC Best Paper Award.
\end{IEEEbiography}

\begin{IEEEbiography} [\addphoto{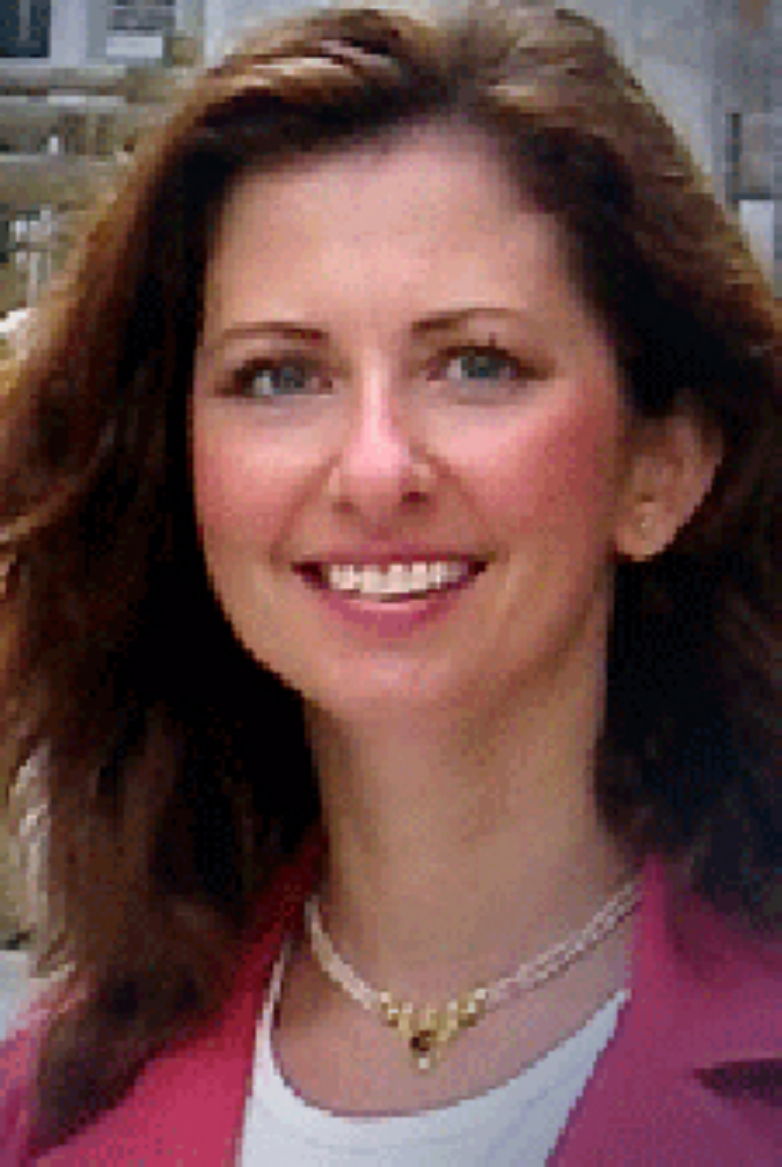}]
{Anna Scaglione}(SM'08, F'11) received her M.S. and Ph.D. degrees in Electrical Engineering from the University of Rome ``La Sapienza", Italy, in 1995 and 1999, respectively. She is currently Professor in Electrical Engineering at the University of California at Davis, CA.  Prior to this she was postdoctoral researcher at the University of Minnesota in 1999-2000, Assistant Professor at the University of New Mexico in 2000-2001, Assistant and Associate Professor at Cornell University in 2001-2006 and 2006-2008, respectively. She received the 2000 IEEE Signal Processing Transactions Best Paper Award, the NSF Career Award in 2002, the Fred Ellersick Award for the best unclassified paper in MILCOM 2005, and the 2005 Best paper for Young Authors of the Taiwan IEEE Comsoc/Information Theory section. Her research is in the broad area of signal processing for communication systems. Her current research focuses on cooperative communication systems and decentralized processing for sensor networks.
\end{IEEEbiography}

\begin{IEEEbiography} [\addphoto{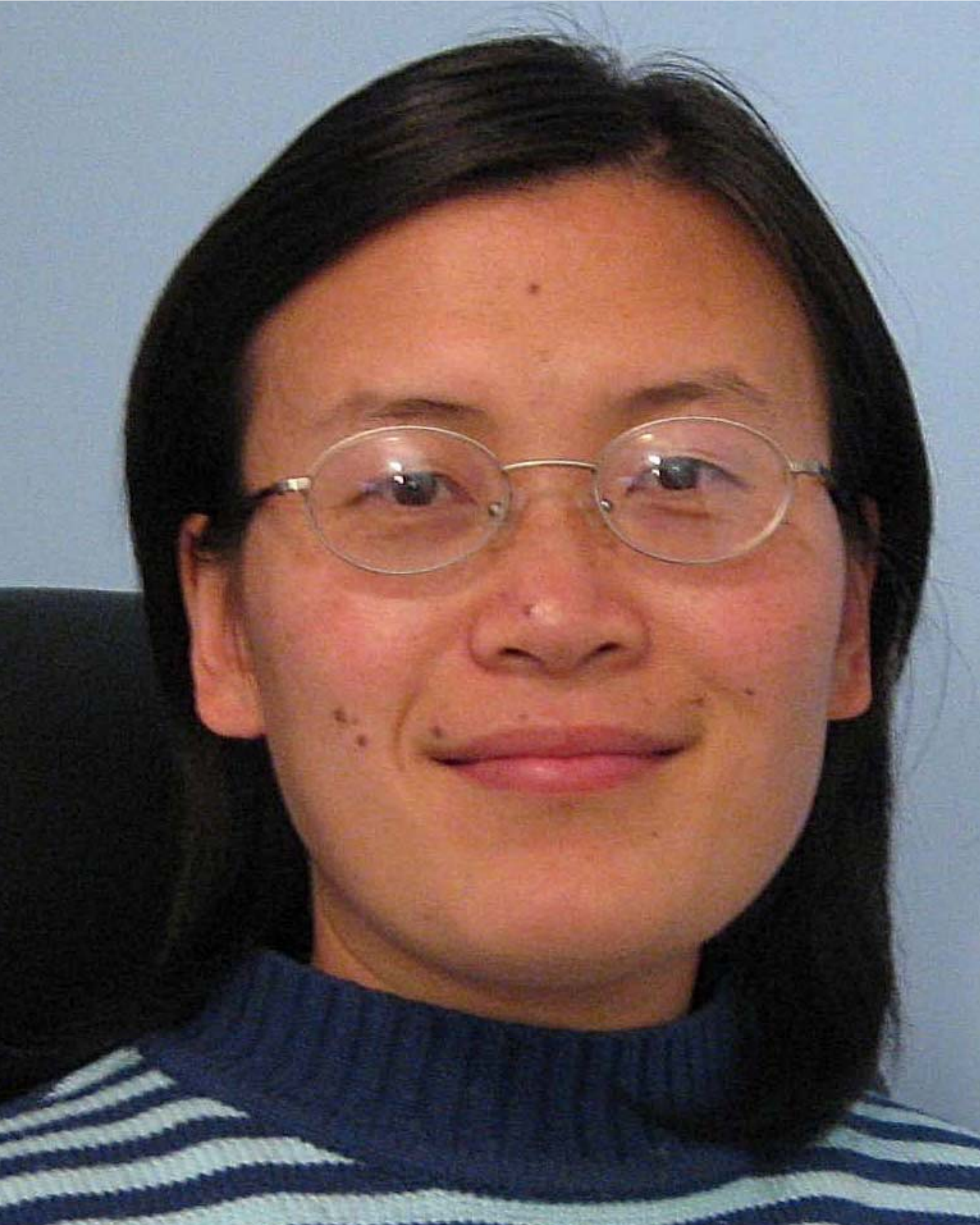}]
{Zhifang Wang} (S'02, M'05) received the B.S. and the M.S. degree in 1995 and 1998 respectively from EE Tsinghua University Beijing, China, and the Ph.D degree in 2005 from ECE Cornell University. She is currently a Postdoc Research Associate at the Information Trust Institute at University of Illinois at Urbana-Champaign. Her research interests include Smart Grid communications, system analysis, cascading failures, and real-time controls of electric power grids.
\end{IEEEbiography}

\balance

\end{document}